\documentclass[fleqn,usenatbib]{mnras}

\usepackage[T1]{fontenc}

\DeclareRobustCommand{\VAN}[3]{#2}
\let\VANthebibliography\thebibliography
\def\thebibliography{\DeclareRobustCommand{\VAN}[3]{##3}\VANthebibliography}

\usepackage{graphicx}	
\usepackage{amsmath}	
\usepackage{txfonts}
\usepackage[usenames,dvipsnames]{color}

\newcommand{\intd}[1]{\;\text{d}#1}
\newcommand{\dx}{\intd{x}}
\newcommand{\dr}{\intd{r}}
\newcommand{\xsurf}{x_{{\text{surf}}}}
\newcommand{\xs}{x_{\text{s}}}
\newcommand{\xt}{x_{\text{t}}}

\newcommand{\rd}{\text{d}}
\newcommand{\ba}{\mathbf{a}}

\newcommand{\covr}{\psi}
\newcommand{\invce}{e}

\newcommand{\omegaac}{\omega_{\text{ac}}}

\newcommand{\Msun}{\mbox{M}_{\sun}}
\newcommand{\GMsun}{G\Msun}
\newcommand{\Rsun}{\mbox{R}_{\sun}}
\newcommand{\Rf}{\mbox{R}_{\text{f}}}
\newcommand{\Rp}{\mbox{R}_{\text{p}}}
\newcommand{\Rph}{\mbox{R}_{\text{ph}}}
\newcommand{\Rphr}{\mbox{R}_{\text{ph,r}}}
\newcommand{\Rs}{\mbox{R}_{\text{s}}}
\newcommand{\Rc}{\mbox{R}_{{\text{ac}}}}
\newcommand{\Rcr}{\mbox{R}_{{\text{ac,r}}}}

\newcommand{\Rr}{\mbox{R}_{\text{r}}}
\newcommand{\Rt}{\mbox{R}_{\text{t}}}
\newcommand{\Rtz}{\mbox{R}_{\text{t},0}}

\newcommand{\Mr}{\mbox{M}_{\text{r}}}
\newcommand{\Mt}{\mbox{M}_{\text{t}}}

\newcommand{\Rpval}{695.78\pm 0.16~\text{Mm}}

\newcommand{\numrestsecs}{8} 

\title{The acoustic size of the Sun}

\author[M.~Takata and D.~O.~Gough]
       {M.~Takata$^{1}$%
\thanks{E-mail: takata@astron.s.u-tokyo.ac.jp (MT); douglas@ast.cam.ac.uk (DOG)}
        and D.~O.~Gough$^{2,3}$\footnotemark[1]\\
        $^{1}$Department of Astronomy, School of Science, 
        The University of Tokyo, 
        7--3--1 Hongo,
        Bunkyo-ku, Tokyo 113--0033, Japan\\
        $^{2}$Institute of Astronomy, Madingley Road, Cambridge CB3 0HA\\
	$^{3}$Department of Applied Mathematics and Theoretical Physics,
        Wilberforce Road, Cambridge CB3 0WA}

\date{Accepted 2023 October 16. Received 2023 October 16; in original form 2023 August 11}

\pubyear{2023}

\begin{document}
\label{firstpage}
\pagerange{\pageref{firstpage}--\pageref{lastpage}}
\maketitle

\begin{abstract}
Analysis of
f-mode frequencies 
has provided a measure of the radius of the Sun which is 
lower,
 by a few
 {hundredths}
 per cent, than
the photospheric radius determined by direct {optical} measurement.
Part of this difference can be understood by 
{recognizing} that it is primarily the 
{variation of density 
well beneath the photosphere
of the star that determines the structure of 
these essentially adiabatic}  
oscillation modes, 
{not some aspect of radiative intensity.} 
In this paper 
we attempt to shed further light on the 
{matter, by considering a differently defined, and dynamically more robust, seismic} radius,  
 namely 
{one determined from}
p-mode frequencies.
{
This radius is calibrated by
the distance from the centre {of the Sun }
to the position in the subphotospheric layers
where
the 
{first derivative of the}
density scale height changes
essentially discontinuously.
}
{We find} that
that radius 
is {more-or-less} consistent
with what is suggested by the
f modes.
In addition,
the interpretation of the radius 
 inferred from
 p modes
 leads us to
{
understand {more deeply} 
the role of
the total mass constraint
in the structure inversions.
This enables us to reinterpret
the 
{sound-speed}
inversion{,}
suggesting that
the positions of the photosphere and
the adiabatically stratified layers in the convective envelope
differ nonhomologously from those of the standard solar model.
}
\end{abstract}




\section{Introduction}\label{sec:introduction}

\begin{table*}
\begin{minipage}{154mm}
\caption{Global quantities concerning the structure of the Sun
and their relative errors.
{
Here,
$\Rph$
means
the photospheric radius,
and
$\Rf$
is
the f-scaled radius
(subject to estimated systematic errors),
as discussed in the text.
The last row presents the main result
of this paper,
the p-scaled radius,
$\Rp$ (quoted with 3-$\sigma$ statistical errors).
}
}
\label{tab:global_quantities}
 \begin{tabular}{clcl}
 \hline
\multicolumn{1}{c}{Quantity} & 
\multicolumn{1}{c}{Value} & 
Relative error & 
\multicolumn{1}{c}{Reference}\\
 \hline
  $\GMsun$ &
  $(1.32712440041 \pm 0.0000000001) \times 10^{26}$~cm$^3$ s$^{-2}$
      & $8 \times 10^{-11}$
      & \citet{Folkner:2009aa}\\
  
 \hline
  $G$ &
       $(6.67408 \pm 0.00031)\times 10^{-8}$~dyn~cm$^2$~g$^{-2}$
      &
      $5 \times 10^{-5}$
      &
      \cite{Mohr:2016aa}\\
 \hline
  $\Rph$ &
      $695.99 \pm 0.07$~Mm & $1\times 10^{-4}$ & \citet{Allen:1973aa}
\\

             &
      $695.508\pm 0.026$~Mm & $4 \times 10^{-5}$ & \cite{Brown:1998aa}
	      \\
  &
      $695.658 \pm 0.140$~Mm & $2 \times 10^{-4}$ &
	      \citet{Haberreiter:2008aa}\\
 \hline
 $\Rf$ &
      $695.68 \pm 0.03$~Mm & $4\times 10^{-5}$ & \citet{Schou:1997aa}
              \\
             &
             $695.787$~Mm & -- &
     \citet{Antia:1998aa}
     \\
 \hline
  $\Rp$
   &
      $\Rpval$ & $2\times 10^{-4}$ & this work
              \\
 \hline
 \end{tabular}
\end{minipage}
\end{table*}

Solar oscillation frequencies have been measured very accurately 
from observations 
{from} space missions and
 ground-based networks. 
The {estimated} relative errors in some of the frequencies 
are now as low as 
the order of $10^{-6}$.
The high-precision data enable us to perform inverse calculations for 
the Sun's seismically accessible
structure variables, such as 
{sound speed}
and density. 
These are commonly accomplished by characterizing in some way the differences between the Sun and a theoretical reference model, which are usually presumed to be small enough for linearization to be valid.  
We note that 
the smaller are the observational errors,  
the larger is the number of the properties that we should take into account
when carrying out the inversions.
An example 
is the error in the solar radius,  which 
is the principal issue
to which this paper is addressed.

For structure inversions it is conventionally assumed 
that
the total radius of the reference model is exactly equal to that of the
Sun.
This assumption should generate no significant error 
as long as the uncertainties in the eigenfrequencies are much larger
than those in the solar radius.
However, 
that is {no longer}
the case.
%
An example is a measure of the solar radius derived from an analysis 
of f-mode frequencies 
from the Solar Oscillations Investigation (SOI)/%
Michelson Doppler Imager (MDI) 
instrument
\citep{Scherrer:1995aa},
on the {\it Solar and Heliospheric Observatory} (SOHO) space mission, 
%
{obtained} 
by \citet{Schou:1997aa};   
{their procedure was to scale the f-mode frequencies of the standard 
solar model S of \citet{Christensen-Dalsgaard:1996aa} to the observations  
{by adjusting $R$ in the approximate} asymptotic f-mode dispersion relation
$\omega^2=\sqrt{l\left(l+1\right)}
{\GMsun}
/R^3${,
where $\omega$, $l$, $G$ and $\Msun$ mean
the angular eigenfrequency of the mode,
the spherical degree of the mode,
the gravitational constant
and
the total mass of the Sun, respectively.}
The resulting `photospheric' radius {turned out to be} about $0.3$~Mm 
lower than the conventional {photospheric}  value, $695.99$~Mm \citep{Allen:1973aa}, of the time; 
{it} has recently been adopted as an IAU standard unit, 
{the nominal solar conversion
 constant for the radius:  ${\cal R}^{\text{N}}_{\sun} = 695.7$~\text{Mm}}
 \citep{2016AJ....152...41P}.}
Here we call it the solar 
`f-scaled radius', 
{$\Rf$}.
\citet{Antia:1998aa}
has also determined an {f-scaled} radius, using 
frequencies obtained from the GONG network 
\citep{Harvey:1996aa}, 
finding it to be lower {than the conventional value} 
 by $0.03$~{per cent}, 
which corresponds to $0.2$~Mm.
\citet{Basu:1998aa}
{then} demonstrated how these apparently tiny differences 
indeed have
considerable influence on inversions
for sound speed;  
although 
the surface helium abundance and the depth of the convection zone
are hardly affected.
%

\citet{Brown:1998aa}
have revised the photospheric radius, obtaining it almost directly 
{from the location of the  inflexion point in the limb intensity;  they used  
two different model atmospheres to estimate the height of the 
inflexion point  above the photosphere, with an essentially 
consistent outcome of $0.5$~Mm.}  
Their result, 
which has been adopted by
\citet{Cox:2000aa},
is 
smaller, by $0.07$~{per cent}
($0.5$~Mm),
than
the earlier value quoted by \citet{Allen:1973aa}
(see {Table} \ref{tab:global_quantities}), 
and is even smaller than the 
{f-scaled} radii   
inferred
by \citet{Schou:1997aa} and \citet{Antia:1998aa}.
{Subsequently, \citet{Haberreiter:2008aa} also estimated the 
inflexion-point height using yet another
model atmosphere. They concluded 
that it is only $0.33$~Mm above the photosphere, leading to a 
photospheric radius closer to the f-scaled radius determined by 
\citet{Schou:1997aa}.
They claimed that their numerical coincidence reconciles 
the discrepancy between the
f-scaled radius and
the conventional value of the radius
quoted by  \citet{Allen:1973aa}.  However, 
{the absence of} an explanation of why
the earlier and essentially identical (albeit with a different model atmosphere)  comparison by \citet{Brown:1998aa} 
led to a different result must surely render that claim premature.}

{Evidently, analyses are now sufficiently precise to exhibit significant differences between radii determined from different structural properties.  Further progress therefore requires appropriate 
distinctions to be made.  The photospheric radius, however defined,}
depends on {equilibrium solar} models that 
involve radiative transfer, which is not 
directly accessible to seismic probing; 
{the f-scaled radius is at least an adjustment that depends on a dynamically 
pertinent property of the hydrostatic structure.}
Additionally, one can define an f-mode radius as that which renders the approximate dispersion 
relation quoted above almost exact. 
{It is essentially
the distance from the centre
to the position of the peak in
the distribution of
kinetic energy density
of each f mode.}
That depends on the stratification of only dynamically pertinent  
variables, principally density \citep{Gough:1993aa}; {it} is a 
weakly increasing function  
of degree $l$.
It is noteworthy that these f-mode radii are well below the 
subphotospheric 
superadiabatic boundary layer, which is located about
{$0.08$}~Mm below
the photosphere in model S.
{In fact, we can estimate 
the f-mode radii 
{directly from the density stratification} of model S
to be about $11$~Mm and $4$~Mm
below the photosphere for $l=100$ and $300$, respectively.
}
{
 One might have suspected it to be possible to identify a unique 
limiting f-mode radius by attempting to extrapolate to infinite 
$l$ an extended asymptotic relation 
\citep[e.g.][]{Gough:1993aa,Dziembowski:2001aa}
accounting for the diminishing vertical extent of the dominating dynamics, at least if density were to vanish at the surface, as it does in a polytrope.   However, although viewed from the interior   the true structure appears to approach such a vanishing situation, located at what we might call a phantom surface (analagous to the phantom acoustic singularity which we address at the beginning of the Section \ref{sec:seismic_radii}), its true behaviour is to extend beyond that surface into the outer atmosphere where the 
acceleration due to gravity
decreases and where the energy density of the mode might even increase, causing the apparent limiting radius to vary with the range of $l$ adopted for the extrapolation 
\citep[e.g.][]{csrdog1994ApJ...423..488R,csrjcd1995MNRAS.276.1003R}.
Moreover, fluid motion associated with turbulence or other 
high-degree oscillations is likely to destroy horizontal 
coherence.}

Solar-cycle variation 
in the 
{f-scaled} 
radius has also been studied  
\citep{Antia:2000aa,Dziembowski:1998aa,Dziembowski:2000aa,Dziembowski:2001aa,Antia:2003aa};
the results are controversial, 
as are 
reported
variations in 
the direct measurement of the photospheric radius
\citep[cf.][]{Gough:2001aa}.
\cite{Lefebvre:2005aa}
and
\cite{Lefebvre:2007aa}
actually inverted 
the temporal change of the f-mode frequencies
to detect nonhomologous
solar-cycle variation
in the structure of the subsurface layers
down to about $97$~{per cent} of the photospheric radius.
It should be noted, however, that
they
neglected the contribution
of the near-surface effect of
turbulence and magnetic fields
to the observed f-mode frequencies
(as \cite{Lefebvre:2007aa} admit explicitly).
This assumption needs to be checked carefully
in further studies.

\citet{Schou:1997aa} {also} {point out that}
we should bear in mind
the possibility that
the f-mode frequencies
could be significantly influenced by
unaccounted processes in
the superadiabatic layer of the convection zone, such as {are produced by} turbulence and the presence of a magnetic field, 
which are notoriously ill understood.
{They} might be 
more susceptible to such processes than are the p modes 
on account of their smaller inertiae,
although they are certainly less sensitive than p modes to the mean stratification 
because they are very nearly uncompressed  \citep{Gough:1993aa}.
\cite{Dziembowski:2001aa} take some account of the surface effects
on  f-mode frequencies by assuming a 
particular dependence on  frequency and mode inertia.
It is not obvious whether the dependence can be justified
from a physical point of view.
In any case,
it is true that
p modes react differently from f modes
to the processes in the convective boundary layer.
We are therefore motivated to determine a solar radius 
from p-mode frequencies {alone}, 
for that is directly pertinent to seismic inferences of the interior structure 
obtained from p-mode inversions.
We show in this paper
that that is possible.

{
\citet{Richard:1998aa} extended
a formula for structure inversion
to take account of the difference
between the radii of
the Sun and the reference model.
By radius they mean the distance between the centre
and the temperature minimum.
The main aim of their study was
to constrain the helium abundance more 
{tightly};
in so doing
they obtained estimates of radius difference
that were so sensitive to the mode sets used
that they concluded it is impossible
to determine the radius of the Sun
at the $10^{-4}$ accuracy level from p-mode frequencies.
The present study aims to
re-examine this conclusion.
}

In view of the increasing accuracy of our endeavour, it behoves us to define 
more precisely what measure of a solar radius we seek to determine.  Here
we adopt a purely seismic definition, which, unlike the photosphere or some 
other thermal structure of the atmosphere, such as temperature minimum, 
can be obtained purely dynamically, and in principle is independent of our
theoretical reference solar model.   We offer, in subsection \ref{015032_29Oct18},
various operational options, whose merits we then discuss.

Another {aspect} of the present study
{is the recognition that}
the
relative errors
in the 
radius 
of the Sun, 
and of the gravitational constant,
have {hardly ever} been 
taken into account
in the integral constraints relating oscillation frequencies to the {Sun's} seismic structure, 
even though
these errors are no longer negligible
compared with the errors in the frequencies.
\citet{Brown:1998aa}
and \citet{Haberreiter:2008aa}
have recently 
revised the estimates of  
the relative error in the Sun's photospheric radius
to $4\times10^{-5}$
and $2\times 10^{-4}$, respectively, 
as recorded in {Table} \ref{tab:global_quantities}.   Moreover, 
the relative uncertainty in {the} gravitational constant, which is
one of the most poorly measured physical constants,
is on the order of
$10^{-5}$. 
Therefore
we have to 
revise the inversion procedure 
so that 
it treats
both the errors in the oscillation frequencies
and those in global quantities consistently.
Fortunately, we need not worry about any uncertainty in
the Sun's mass $\Msun$ 
{per se}, 
because it is always 
multiplied by
the gravitational constant $G$
in the equations of helioseismic inversion
(see {Section} \ref{sec:formulation}); 
the quantity $\GMsun$ has much smaller errors than 
any of the observed oscillation frequencies,
{and is indeed} one of
the best determined, 
by radar-echo measurements of the planetary motion
(cf.~{Table}~\ref{tab:global_quantities}),  constants in astronomy.

{As we have already announced,} the primary purpose of this paper is to
formulate a well defined method to estimate a radius  of the Sun 
using p-mode frequencies.  {As we shall see, this requires including the requirement 
that the total mass $M$ (actually $GM$) agrees with observation.} 
We discuss the physical meaning of this radius, which we call the {p-scaled} radius,  $\Rp$,
in the light of the f-{scaled} radius $\Rf$. 
The secondary purpose is 
to revise the inversions for the structure of the Sun
by taking account of both the differences and the uncertainties in what one might call the  
total radius $R$ {(in a sense that we shall propose later)} 
and the gravitational constant $G$.
To this end, we need to extend the inversion formulae so that they
take account of these differences{:} the total radius $R$ and the gravitational constant $G$
(and,  for completeness, the product $GM$ of the gravitational constant and the total mass).
This not only allows us to revise the structure inversions, 
but also enables us to perform a direct inversion for the
difference  $\delta R$ between the total radii of the Sun and the reference model.
{Properly executed, the outcome provides a measure of a solar radius 
that is independent of any reference theoretical model.}
%
{
That property is a property of only the dynamically pertinent variables, which
are not directly accessible to astronomical observation.  Accordingly, we relate
$\Rp$ to the photospheric radius $\Rph$, whose value depends on
radiative transfer, which is not itself dynamically relevant.  That necessarily
involves comparison with a theoretical solar model.  We adopt 
expression{s 
(derived in {Sections} 
\ref{015032_29Oct18} 
{and \ref{subsec:rel_to_acr}}) to identify $\Rp$, 
which}  
{is obtained by scaling} our reference model, namely {model} S. 
}

The rest of this paper consists of
\numrestsecs\ 
sections.
{
We first present 
in {Section} \ref{sec:seismic_radii}
physical pictures about
seismic radii.
}
In {Section} \ref{sec:formulation}, 
we 
extend the formulae for the structure inversions
to include the effect of the
differences in
the radius $R$ and the gravitational constant $G$.
In {Section} \ref{sec:R_inversion},
we
propose a method to infer the radius difference
based on the p-mode frequencies, and
give an interpretation of
the 
{p-scaled}
radius, {$\Rp$}; 
{Section} \ref{sec:revision}
describes 
an accompanying inversion procedure
for the sound-speed and density structures.
In {Section}
\ref{sec:numerical_tests} we examine how well the prescribed methods work
based on the known structures of theoretical models, 
%
{and}  we estimate the 
{p-scaled} radius
using the real data
in {Section} \ref{sec:real_R_inv}.
%
%
Section
\ref{sec:discussion} { is a discussion of our proposed procedure, 
and our conclusions are summarized } {briefly} 
in {Section} \ref{sec:conclusion}.
Preliminary results of this paper have been 
reported by
\citet{Takata:2001aa,Takata:2003ab}.

Before proceeding, we make the obvious remark that without a well defined procedure for 
characterizing the Sun's radius in terms of its {seismologically} accessible structure, its {unambiguous} 
value cannot be determined by seismology alone.  However, it is possible to determine a 
radius change by scaling the independent variable used in a reference model to produce 
a representation that is in some sense close to the target structure.  As we 
explain below, that is what our inversion procedure achieves 
automatically.  Here we wish to relate the radius to the 
structure throughout the deep interior, rather than just to the 
near-surface 
layers which are severely susceptible to uncertainties in the physics.  This 
is likely 
to be more robust from p modes than from f modes. We 
{demonstrate} below how that is accomplished.

\section{Seismic radii}
\label{sec:seismic_radii}

\begin{figure}
\centering\includegraphics[width=\columnwidth]
{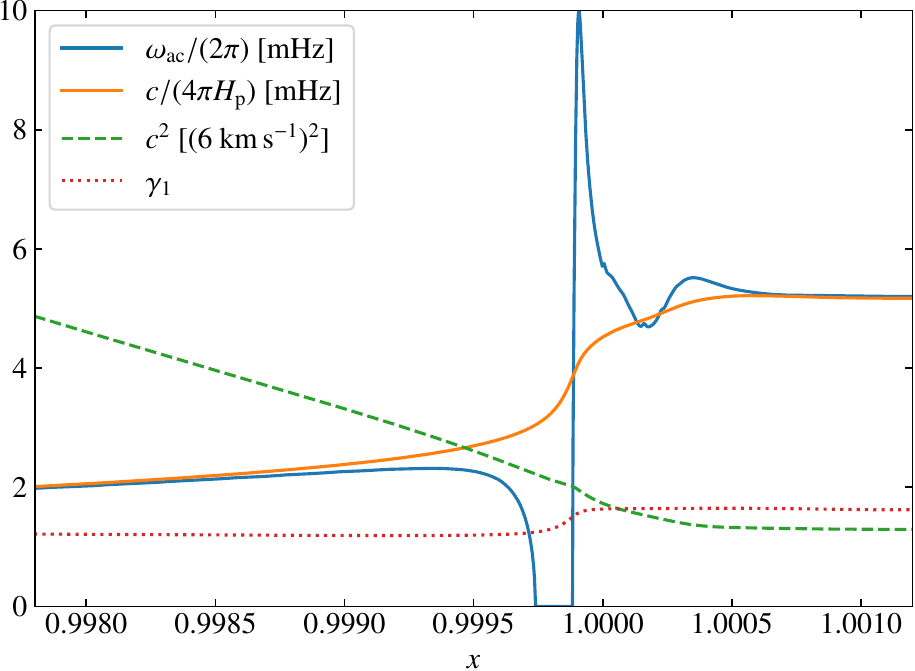}
\caption{
{
Acoustic cut-off frequency $\omegaac$ 
associated with the Lagrangian pressure-perturbation 
eigenfunction,
defined by
equation
(\ref{eq:acutoff}),
in the surface layers 
of 
solar model S of 
\citet{Christensen-Dalsgaard:1996aa}, 
 plotted as a continuous {blue} curve against
fractional radius $x=r/\Rph$,
where 
$\Rph$ stands for the photospheric radius.
Between about $x=0.99974$ and $x=0.99988$, $\omegaac$ is 
imaginary, and formally wave propagation can occur at all frequencies. 
The continuous orange
curve is the corresponding cut-off frequency 
adjusted according to 
 the procedure proposed by \citet{Langer:1937aa} to prevent the error in the JWKB approximation to the wave equation (which experiences an appropriately smoothed background state) from approaching divergence at the phantom singular point at which pressure would vanish, leading to $\omegaac \simeq c/(2H_{\text{p}})$, where $H_{\text{p}}$ is the pressure scale height. 
 The {dashed green} curve is the {squared} adiabatic
 {sound-speed}
 (in units of {$(6~\text{km}~\text{s}^{-1})^2$}) and the {dotted red} curve 
is $\gamma_1$.
{The almost linear section 
{of the squared sound-speed}
below $x \approx 1.0000$ could in principle be extrapolated to zero to locate a phantom singularity, as indicated by 
 \citet{Balmforth:1990aa};  however the deviation from linearity renders the outcome insufficiently precise for our purposes here.} 
The top of the superadiabatically stratified convection zone
is located at $r/\Rph \approx 0.99994$.} }
\label{fig:cutoff_freq}
\end{figure}

{
When we embarked on our investigation we thought to provide a well defined
dynamically pertinent radius of the Sun, determined from p-mode frequencies.  
The f modes were ignored because they are relatively more concentrated near
the surface, and are likely to be more influenced by the vagaries of the turbulence 
in the upper boundary layer of the convection zone.  Noting that near the surface 
{the squared sound-speed}
declines almost linearly with radius \citep[cf.][]{Balmforth:1990aa}, as it
does also in a complete plane-parallel polytrope whose pressure and density 
vanish together at its surface, 
we had in mind representing the Sun by a near-polytropic analogue whose effective 
surface can be determined purely dynamically.  As has been evident since early 
studies of atmospheric oscillations \citep[e.g.][]{lamb1911RSPSA..84..551L,lamb1932hydrodynamics}, the polytropic surface is a 
singular point of the governing dynamical equations. 
In reality, the solar envelope does not 
resemble a polytrope as far out as the latter's surface, but undergoes a transition 
to almost isothermal stratification near the photosphere.  However, because what
the oscillations experience is mainly the form of the declining pressure and density 
beneath their upper turning point, where 
{$\omegaac = \omega$}, they 
behave as though a phantom singularity actually exists.  Therefore, provided the 
mode frequencies are considerably lower than the acoustic {cut-off},
the near-isothermal 
atmosphere is of lesser dynamical import, as has been demonstrated explicitly 
by \citet{Christensen-Dalsgaard:1980aa}.  
Consequently we had in mind 
defining an acoustic solar surface as the location of either the phantom singularity,
which we call $\Rs$, or the location $\Rc$ of the region of
extremely rapid variation of $\omegaac$ (see {Fig.~}\ref{fig:cutoff_freq}),
which essentially identifies the upper turning points.  However, 
although it is possible to establish a mathematical procedure to determine such  
location{s}, the outcome does not relate in a straightforward  way to 
what astrophysicists can find useful.  So instead we have decided simply to scale 
the reference model by {a} factor 
{determined from an} integral relation 
{(derived in {Section} \ref{015032_29Oct18}) for 
{$\Rc$}}, and then 
adopt the resulting photospheric radius
{as $\Rp$,}
namely where the 
matter temperature equals the effective temperature.  
{In other words,
we assume
\begin{equation}
\frac{\Rp - \Rphr}{\Rphr}
= \frac{\delta\Rc}{\Rcr}
\;,
\label{eq:scaling_assumption}
\end{equation}
where {the}
subscript r {denotes} quantities
{pertaining to} the reference model, and
$\delta$ means the difference between
the Sun and the reference model.}
Our result is therefore 
not strictly based on dynamics alone.  
But since the phantom {acoustic} surface 
and the turning surface, especially the latter,
are very close to the photosphere,
%
%
any error in the thermal stratification of the reference induces an
error {hardly} greater than that of our derived scaling factor. 
}
{The meanings of 
{the photospheric radius and}
the seismic radii, which are 
introduced in Sections
\ref{sec:introduction} and \ref{sec:seismic_radii},
are summarized in Table~\protect\ref{table:seismic_radii}.}
\begin{table}
\caption{{Meanings of {various} radii.}}
\label{table:seismic_radii}
{%
\begin{tabular}{ccp{46mm}}
\hline
{radius} & & {meaning}
\\
\hline
{photospheric radius} & {$\Rph$} &
{%
distance from the solar centre to
the photosphere, which is characterized as
{the} layer {at} 
optical depth {unity} 
at a particular (visible) wavelength: a 
commonly used {value is} 
500 nm.
{Alternatively, i}f the atmosphere is in
local thermodynamic 
equilibrium,
the photosphere is the layer
where the local temperature is equal to
the effective temperature.}
\\
\hline
seismic radius & & meaning
\\
\hline
f-mode radius & &
distance from the {solar} centre to 
{the centre of energy of each f-mode (essentially}
the peak
in the kinetic-energy distribution{)}
\\
f-scaled radius & $\Rf$ &
photospheric radius calibrated by 
f-mode radii
\\
phantom singularity
& $\Rs$ &
distance from the centre to
the {apparent} zero point of the squared sound-speed
that can be located
(above the photosphere)
by extrapolating the almost linear distribution
in the adiabatically stratified layers
of the convective envelope
\\
acoustic radius & $\Rc$ &
distance from the centre to the subphotospheric
layer
where the 
acoustic cut-off frequency changes extremely rapidly
(essential{ly a} discontinuity {in
the 
{first}
derivative} of the density scale height)
\\
p-scaled radius &
$\Rp$ & 
photospheric radius calibrated by 
$\Rc$
\\
\hline
\end{tabular}}
\end{table}

If both the target structure and the reference model
have similar density and sound-speed profiles
near their upper turning points,
the 
{p-scaled}
radius difference could be identified
with the difference in the position of
the upper turning points themselves.
The upper-tuning points of
p modes
are determined by
an 
acoustic cut-off frequency,
one of which, when Lagrangian pressure perturbation is adopted for describing 
the mode, is given
in terms of the {adiabatic} 
{sound-speed}
$c${,} 
the density scale height $H$
{and the absolute radius
$r$ (i.e, the distance from the centre)
}
by
\begin{equation}
 \omegaac
{:=}
 \frac{c}{2H}
\left(
1 - 2 \frac{\rd H}{\rd r}
\right)^{\frac{1}{2}}
\label{eq:acutoff}
\end{equation}
\citep[e.g.][]{Deubner:1984aa,Christensen-Dalsgaard:1991aa}.

{{The u}pper turning point {of a} radial mode 
{is}  
located where 
$\omegaac$ is equal to the 
{angular}
frequency $\omega=2\pi\nu_{n,0}$ 
of the mode.   
{Here, $\nu_{n,0}$ is the cyclic frequency
of the radial mode with radial order $n$.}
{It is where propagation gives way to evanescence.  It}
is also 
the approximate {turning} location for 
nonradial modes, except when {the degree}  $l$ 
is very large \citep{Gough:1993aa}.}
Other representations of 
{the acoustic cut-off frequency} 
corresponding to other 
dependent (and  independent{)} variables
are similar, and are {barely} distinguishable within the context 
of wave reflection. 
The value  
{pertaining to the Lagrangian pressure perturbation,} given by equation (\ref{eq:acutoff}), {is}
plotted in 
Fig.~\ref{fig:cutoff_freq}.
Because,
as in the realistic structure of the Sun,
scale heights 
vary extremely rapidly
in a thin layer
between the uppermost part of the convection zone
and the base of the isothermal atmosphere,
the acoustic cut-off frequencies exhibit a
sharp, almost discontinuous, incline, 
{presenting an effective wall obstructing further outward propagation.}
Since
it is this wall
which
determines the upper turning points of 
the high-frequency p-modes observed in the Sun,
the distance
between the {solar} centre
and
this wall
should essentially be regarded
as 
{the radius, $\Rc$, {out to which} p modes probe. 
{To be more precise, we can define $\Rc$ to be the location 
of the maximum of $|\rd^2H/\rd r^2|$}, 
{where $\rd H/\rd r$ changes 
almost discontinuously,}
and that can be used to calibrate
the p-scaled radius, $\Rp$}.
{We call $\Rc$ the acoustic radius.%
\footnote{{Although
the term ``acoustic radius'' is widely accepted
in helioseismology
to mean the acoustic travel time
across the solar radius,
there should be no confusion {here} with the
present definition, which has dimension of length.}}
Although,
from a physical point of view,
$\Rs$ can be also regarded
as another kind of acoustic radius,
we choose to attribute the name to only
$\Rc$ in this paper,
because it plays a {greater} role than $\Rs$ does.
}
{
{It is}
a representative value of
the upper turning points of p modes,
the position{s} of 
}  
{which} 
{are} increasing function{s} of mode frequency. 
However, 
{the} rapid variation in $\omegaac$ with height 
near the surface
{renders}  
the 
{
turning points
}
essentially independent
of the mode set used in the analysis{.}  
In the example shown in Fig. \ref{fig:cutoff_freq},
p modes of frequency higher than 
about $2.3$ mHz (and below $5.2$ mHz)
have almost the same upper turning point
at $r/\Rph \approx 0.9999$.
Therefore
any mode set that includes some of such high-frequency modes
should give the same 
{p-scaled}
radius.

According to the above interpretation, 
we may say that
the 
{acoustic}
radius,
$\Rc$,
and
the photospheric radius
share their origin
because
the rapid change in the density scale height $H$ 
(and its derivative)
near the top of the convection zone
is accompanied by
the corresponding change in the optical depth.
In fact, 
the position of
the vertical wall in Fig. \ref{fig:cutoff_freq}
approximately corresponds to
the base of
the photospheric layer
in 
{a typical} atmospheric model of the Sun
\citep[cf.][]{Cox:2000aa}.
{That is much higher than the position of
the peak 
in the distribution
of the kinetic energy density
of typical f modes.}
Therefore
$\Rc$
is a better probe
of
the photospheric radius of the Sun
than is $\Rf$.

\section{%
The principles of the structure inversion,  
and the radius difference
}
\label{sec:formulation}

\subsection{Inversion formulae
that take account of radius and mass differences}

\subsubsection{Relation between the frequency difference
and the structure difference}

Typically, helioseismological inversion aims at obtaining a  representation 
of the difference in structure between the Sun  (or a target theoretical model) 
and a reference model that is believed to be sufficiently close to the target 
for linearization to be valid. 
The small differences
in the structure variables 
are derived from 
integral equations which
relate the structure differences 
to the eigenfrequency differences.
The weighting functions 
in the integrands  that 
multiply the differences (either absolute or relative) in the structure variables 
are called data kernels; they 
depend on both the eigenfunctions and the equilibrium structure, and 
are obtained by perturbing the full 
integral expressions for the eigenfrequencies. 
In principle the integration should extend over the entire domain occupied by 
the star: formally that is to infinity,  but in practice it is adequate to 
truncate the outer limit to a surface so 
far beyond the region of propagation of the oscillation modes that any 
appropriate boundary 
condition contributes negligibly to the integrals. However, if the target were to 
be another theoretical model, that model might have a genuine surface at which 
pressure, and {probably} sound speed, vanish.  We shall sometimes find it convenient to speak 
in terms of such target models when 
describing the properties of the inversion procedures.  Whether the target 
is such a finite model, or a more realistically extended model or the real Sun, 
the integral representations of the eigenfrequencies essentially (that is, to a fair degree of accuracy) satisfy  a variational 
principle
\citep{LedouxWalraven1958,Chandrasekhar:1964aa},
and the perturbations to the eigenfunctions do not 
appear in the formulae for the kernels.
In  more realistic situations in which the boundary conditions appear to preclude
a variational principle, one can construct appropriate integral expressions for the
frequency differences as a perturbation in terms of the eigenfunctions of only the
reference model \citep[cf.][]{veronis1959JFM.....5..401V}.
That suggests that the outcome of the
mathematical process of inversion that does not take explicit account of a
potential radius difference can provide a correct representation of the internal
structure{, as we discuss below}.  However, it 
{does require} in addition a precise definition of a seismic
radius in order to establish an appropriate scaling of the reference model.

\subsubsection{Optimally localized averages of the structure}

In this paper, we adopt optimally localized averaging (OLA) as the means of 
representing the structure differences.  The averaging kernels are constructed 
as unimodular linear combinations of the data kernels, the coefficients $a_{n,l}$ having 
been determined as a {trade-off} between the degree of localization and the resultant 
magnification of the errors in the frequency data $\nu_{n,l}$; the corresponding combinations of the 
data, $\Sigma_{n,l} a_{n,l} \delta \nu_{n,l}/\nu_{n,l}$, are, to within data errors, averages of the true differences between 
the structures of the target star and the reference model.  Such combinations were used originally 
in geophysics just to assess the resolving power of {Whole-Earth} data 
\citep{Backus:1968aa}.
But here we use the averages to represent the structural
differences themselves, bearing in mind that they are not pointwise values but 
in some sense a smoothed version of the true differences.  Treating them as pointwise values, which has been tempting to some, does not normally lead to functions that satisfy the integral relations from which they were constructed, which is why they were not used by geophysicists, at least in the early days, to represent the functions themselves.  Functions  that
do satisfy the integral relations  can easily be constructed simply by insisting, 
subject to certain {demands} introduced to render the outcome determinate,
that the functions have the correct averages. 
{One way to accomplish that is simply to determine the 
smoothest function that satisfies the averaging  constraints.}
{This is a typical problem in calculus of variations with constraints, which can be solved by any standard method.}
At this stage of our discussion,  for the purposes of appreciating the outcome of 
the inversion it is not necessary to know how the averaging 
coefficients $a_{n,l}$ are determined.   
All that is needed, aside from the averages themselves, is to 
know the kernels over which the structure variables are averaged, and, of course, the uncertainties in those 
averages resulting directly from the uncertainties in the measured oscillation frequencies. 

\subsubsection{Formulae for the frequency difference
taking account of radius and mass differences}

As we mentioned in {Section} \ref{sec:introduction},
in conventional helioseismological inversions
it is assumed that there is no difference between the total 
radii $R$ of the reference
model and of the Sun, nor any error in the gravitational constant $G$.  
%
%
In a similar fashion  
we obtain
for the frequency differences between the Sun and 
the reference model, which constitute the data to be inverted:
\begin{equation}
\frac{\delta\nu_{n,l}} {\nu_{n,l}} =
\int K_{\covr,\rho}^{(n,l)} \frac{\delta_x \covr}{\covr} \dx
+
\int K_{\rho,\covr}^{(n,l)} \frac{\delta_x (G \rho)}{G \rho} \dx
+
S_{n,l}
\;,
\label{eq:perturbed_VP}
\end{equation}
for each mode.
The details of the derivation of this equation are given in
appendix \ref{sec:derive_pt_VP}.
The meanings of the symbols 
in equation (\ref{eq:perturbed_VP})
are as followings:
$\nu_{n,l}$ is the cyclic frequency of 
the mode of {radial} order 
$n$ and {spherical} degree $l$ -- we concentrate on the spherically averaged structure, and 
therefore interpret $\nu_{n,l}$ as the uniformly weighted average over azimuthal 
{order} $m$ of the singlet frequencies $\nu_{n,l,m}$ 
\citep[e.g.][]{Ritzwoller:1991aa}
--
$\covr$ is defined by
\begin{equation}
\covr := \frac{c}{r}
\;{;}
\label{eq:covr_def}
\end{equation}
$\rho$ is density;
$x$ is the fractional radius $r/R$, where $R$ is a fiducial (acoustic) radius of the Sun 
whose meaning we discuss later;
{$\delta_x f$} means the difference between some structural variable $f$
pertaining to the Sun and
to the
reference model
at the same fractional radius $x$; and 
$K_{\covr,\rho}^{(n,l)}$ and $K_{\rho,\covr}^{(n,l)}$ are data kernels 
(Fr\'echet derivatives)
for
$\delta_x \covr/\covr$
and
$\delta_x (G\rho)/(G\rho)$
respectively;
$S_{n,l}$ is called a surface term, and 
is introduced 
to take account of uncertainties
in the near-surface regions of the Sun
\citep*[cf.][]{Dziembowski:1990aa}.
We adopt the notation in this paper that,
when no explicit bounds are indicated
the domain of integration   
ranges over
the entire mass of the structure.

We point out the following important features of
equation (\ref{eq:perturbed_VP}):
\begin{enumerate}
 \item
the expressions for the kernels $K_{\covr,\rho}^{(n,l)}$ and $K_{\rho,\covr}^{(n,l)}$ are
      essentially
      the same as those
      in the conventional structure inversions,
      which are derived under the assumption that there is no difference in the
      global quantities $R$ and $G$
      \citep[e.g.][]{Gough:1991aa}
      ;
 \item  
	the differences 
	$\delta_x \covr/\covr$
	and
	$\delta_x (G\rho)/(G\rho)$
	between the Sun and the reference model are taken
not at the fixed radius $r$
but at the fixed fractional radius $x$, and are regarded as functions of $x$;
 \item
the difference in the fiducial radius $R$ is present implicitly in
the expressions
through the difference {operator} $\delta_x$ at fixed $x$, 
although here
we give no physical meaning to $R$ -- 
at this stage it is just a scale factor -- 
we attempt to
provide a physical interpretation later; 
 \item
density $\rho$ is multiplied by the
gravitational constant $G$ wherever it appears;
 \item
for this particular choice of structure variables, 
namely $\covr$ and $G\rho$,
neither
the difference in the total mass $M$
nor that in $GM$ appears explicitly in the expression;
this is not the case for all choices of structure variables, although the current choice is not unique in this respect (see Appendix \ref{sec:invform}); we emphasize that the total mass constraint (\ref{eq:mass_conservation}) was not incorporated {explicitly} 
into the derivation of these data kernels.  
{Therefore equation (\ref{eq:perturbed_VP}) is applicable to asteroseismology too. 
\citet{Gough:1993ab} have 
demonstrated its use in that situation, where additional, non-seismic,
data to estimate  the radius and mass of the star {are} required.}
%
\end{enumerate}

\subsubsection{Total mass constraint}

The total mass constraint,
\begin{equation}
 M = \int 4\pi r^2 \rho \dr
  \;,
\label{eq:mass_conservation}  
\end{equation}
which we usually adopt to ensure that the inversions are
consistent with the observed value of the solar mass,
can similarly be extended to include
the difference in the total radius $R$ and
the product of
the gravitational constant
and
the total mass; {it may be written:}
\begin{equation}
 \frac{\delta R}{R} = \frac{\delta (GM)}{3GM}
- \int \frac{4\pi R^3 x^2 \rho}{3M} \frac{\delta_x (G\rho)}{G\rho} \dx
\;,
\label{eq:mass_constraint}
\end{equation}
{in which} 
$\delta R$ and $\delta (GM)$ appear explicitly.
Since the form of equation (\ref{eq:mass_constraint}) 
is similar to that of equation (\ref{eq:perturbed_VP}),
it is common {practice} to treat all of these equations in like manner, with no caution
as to the different structural connotation.  
{
On the other hand, in the present analysis,
we explicitly distinguish
equation (\ref{eq:mass_constraint}) from
equation (\ref{eq:perturbed_VP})
based on their physical meanings.
This is essential to discuss the radius difference
between the target structure and the reference model.
}

\subsection{Annihilator relation 
associated with a uniform scaling}
\label{subsec:annihilator_relation}

So far we have offered no insight into the physical meaning of $R$ 
in equations
(\ref{eq:perturbed_VP})
and
(\ref{eq:mass_constraint}).
To assist thinking,
we first draw attention to the following
annihilator relation:
\begin{equation}
 \int K_{\covr,\rho}^{(n,l)} \frac{{\rd}\ln\covr}{{\rd}\ln r} \dx +
 \int K_{\rho,\covr}^{(n,l)} \frac{{\rd}\ln \rho}{{\rd}\ln r} \dx = 0\;,
\label{eq:annihilator}
\end{equation}
for any $n$ and $l$. 
Appendix \ref{sec:derive_ann} provides a proof of this relation,
which is closely associated with 
the
homology relation
that preserves 
adiabatic eigenfrequencies:
if the radial
coordinate is uniformly stretched by a constant factor $\lambda$ according to 
\begin{align}
 r &\rightarrow  \lambda r\;,
\label{eq:homologous_r}
\end{align}
and the profiles of the other {seismologically} accessible variables are scaled as  
\begin{align}
\covr(r) &\rightarrow \covr(\lambda r)
\\
 G \rho (r) &\rightarrow  G \rho ( \lambda r )\;,
\\
 R & \rightarrow \lambda R
\label{eq:homologous_R}
\\
\noalign{\noindent and}
 GM & \rightarrow \lambda^3 GM
\;,
\label{eq:homologous_M}
\end{align}
the eigenfrequencies $\nu_{n,l}$ are unchanged.
{
There is an implicit assumption
in this argument
that $K_{\rho,\covr}^{(n,l)}=0$
at the upper bound{ary} of the integral
in equation (\ref{eq:annihilator}).
}
This {homology relation}
can be regarded
as a generalization of the
popular statement
in the theory of stellar pulsation
that
the eigenfrequency of the radial fundamental pulsation mode of a star is 
proportional to 
$\sqrt{GM/R^3}$.
We should stress that
this statement is only approximately true,
whereas 
the homology relation is exact 
for all linearized adiabatic oscillations.

The homology relation demonstrates the important fact that
there exists a series of (isospectral) structures
that cannot be
distinguished by their eigenfrequencies alone.
Owing to this characteristic,
it is evident that there is an
ambiguity 
of the stretching of the radial coordinate
in any outcome of
inversions that disregard the total mass constraint.
The isospectral structures resulting from only such stretching have different 
masses, so one can isolate an acceptable one according to 
its total mass.  However, there remains a formally infinite set
of
{seismologically} acceptable 
structures, not necessarily with the same radius, whose differences lie in the 
annihilator of the data kernels.

In practice, however,
one obtains results from OLA with hardly an apparent ambiguity, 
whether the mass constraint is included or not.
Therefore
any stretching factor appears to be determined
implicitly in the inversion procedure.
We need to know which specific value is chosen.
For example,
were
{it the case that} 
the target structure and
the reference model
{were actually} related strictly homologously 
in the sense specified by
equations
(\ref{eq:homologous_r})--(\ref{eq:homologous_M}), 
{then} $\delta \nu_{n,l} = 0$ for all modes,
and any linearized inversion      
in which the inferences are expressed by
a linear combination of the frequency differences 
would
result in there being no difference between (at least the localized averages of) 
the target and the
reference.
This means that
the homology factor $\lambda$
in equations (\ref{eq:homologous_r})--(\ref{eq:homologous_M})
is implicitly
detected and
is 
properly related to
the scale-factor difference $\delta R$
in the inversion procedure, 
with 
$\delta_x \covr = 0, \delta_x (G\rho) = 0$
in equation (\ref{eq:perturbed_VP}). 
However, we do not know the value of $\delta R$ at this stage.
This example suggests that
there is some principle that 
determines the scale factor $R$
even if the differences are not homologous.

{Here we make a remark about
the conventional method of 
structure inversion,
which usually assumes no difference
between the radii
of the target structure
and the reference model.
}
We can {validly} 
reinterpret conventional inversions
if they are performed without explicit use of the total mass constraint.
In that case, we should replace
the labels, 
{setting}  
$\delta_r c/c$ and $\delta_r \rho/\rho$  {to} 
$\delta_x \covr/\covr$ and $\delta_x (G\rho)/(G\rho)$,
respectively.
{Then we do not}  
know the radius difference,
which is required for the operator $\delta_x$
to be well defined,
{until we carry out separately} the additional inversion
for $\delta R$
using the total mass constraint.

In summary, we have two questions to answer here:
\begin{enumerate}
 \item
How can we know the stretching factor,
$\lambda = 1 + \delta R/R$, that is implicitly
determined in the procedure of the OLA method?
\label{it:Q1}
 \item
What kind of principle is operative in the process of determining 
an appropriate value of that stretching factor?
\label{it:Q2}
\end{enumerate}
We answer question \ref{it:Q1} immediately in the following section;
question \ref{it:Q2} is addressed in {Section} \ref{subsec:interpretation_of_dRR}.

\subsection{How to determine the radius difference}
\label{subsec:radius_determination}

The total mass constraint (\ref{eq:mass_constraint}),
which
is a physically different condition from
the frequency equation (\ref{eq:perturbed_VP}),
immediately gives us an answer to the first question.
Once we have the density profile
$\delta_x (G\rho)/(G\rho)$ without knowing what $R$, hence $x$, is,
we can substitute this profile
into equation (\ref{eq:mass_constraint})
to get $\delta R/R$.
We could also perform a different 
inversion
for $\delta R/R$ based on equations
(\ref{eq:perturbed_VP})
and
(\ref{eq:mass_constraint}),
which will be described in detail
in {Section} \ref{sec:R_inversion}.
From a physical point of view,
we determine the size of the target star,
which is otherwise ambiguous owing to the undetermined stretching of the
radial coordinate,
by constraining its total mass (actually $GM$).
This answers question \ref{it:Q1}.
We need to know more about the mathematics behind the OLA method
for the structure, which we discuss in
{Section} \ref{sec:R_inversion},
before we can answer question \ref{it:Q2}.

\subsection{OLA and an annihilator vector}

\label{subsec:annihilator}

In the OLA method 
we make inferences about structure variables such as
sound-speed differences 
in the vicinity of some chosen point 
$x=\xt$
by constructing new kernels $\mathcal{K}^{(\covr)}_{\covr,\rho}$ and 
$\mathcal{K}^{(\covr)}_{\rho, \covr}$ as linear combinations of $K_{\covr,\rho}^{(n,l)}$ 
and $K_{\rho,\covr}^{(n,l)}$:
\begin{equation}
\begin{pmatrix}
 \mathcal{K}^{(\covr)}_{\covr,\rho}
 \\
 \mathcal{K}^{(\covr)}_{\rho,\covr}
\end{pmatrix}
:=
 \sum_{n,l} a_{n,l} 
\begin{pmatrix}
K_{\covr,\rho}^{(n,l)}
\\
K_{\rho,\covr}^{(n,l)}
\end{pmatrix}
\, ,
\label{eq:av_c_kernels}
\end{equation}
and define a localized average of
$\delta_x \covr/\covr$
as 
\begin{align}
\left[
\overline{%
\frac{\delta_{\xt} \covr}{\covr}}
\right]_{\text{OLA}}
&:=
\int
 \mathcal{K}^{(\covr)}_{\covr,\rho}(x;{\xt})
\frac{\delta_x \covr}{\covr}(x)
\dx
\nonumber\\&\phantom{=}\mbox{}
+
\int
 \mathcal{K}^{(\covr)}_{\rho, \covr}(x)
\frac{\delta_x (G \rho)}{G \rho}(x)
\dx
\;,
\label{eq:OLA_c_est}
\end{align}
by trying to demand that
the averaging kernel
$\mathcal{K}^{(\covr)}_{\covr,\rho}(x;\xt)$
be 
localized around $x = \xt$
and unimodular 
(i.e. whose integral is unity),
and 
that
the cross-talk kernel
$\mathcal{K}^{(\covr)}_{\rho, \covr}(x)$
is negligibly small everywhere.
The constants $a_{n,l}$ are called inversion coefficients (and are distinct from the components of the annihilator vector $\bf a$ introduced in equation (\ref{eq:annihilator_comp}) below). 
Then
$\sum_{n,l} a_{n,l} \delta \nu_{n,l}/\nu_{n,l}$
estimates the 
localized average of
$\delta_x \covr/\covr$,
somewhat, yet, we hope, not unduly contaminated by 
$\delta_x (G\rho)/(G\rho)$.
As we have pointed out already, there is no need to enquire how the inversion 
coefficients $a_{n,l}$ are determined (although we do sketch a commonly adopted procedure 
in Section \ref{sec:revision}); 
to appreciate the meaning of 
the inversion {all that is necessary} is to know the averaging kernel at each target location $\xt$ and the corresponding cross-talk kernel. 
Similarly, one can attempt to construct a corresponding 
localized average of 
$\delta_x (G\rho)/(G\rho)$
with a localized
unimodular kernel  $\mathcal{K}^{(\rho)}_{\rho,\covr}$ and negligible cross-talk kernel 
$\mathcal{K}^{(\rho)}_{\covr,\rho}$ given by 
\begin{equation}
\begin{pmatrix}
 \mathcal{K}^{(\rho)}_{\covr,\rho}
 \\
 \mathcal{K}^{(\rho)}_{\rho,\covr}
\end{pmatrix}
:=
 \sum_{n,l} b_{n,l} 
\begin{pmatrix}
K_{\covr,\rho}^{(n,l)}
\\
K_{\rho,\covr}^{(n,l)}
\end{pmatrix}
\; .
\label{eq:av_rho_kernels}
\end{equation}
Were the averages to be well localized everywhere,
one could attempt to construct plausible pointwise representations of 
$\delta_x \covr/\covr$ and $\delta_x (G \rho)/(G \rho)$ to estimate the 
cross-talk integrals, and so iterate on the procedure for determining 
the optimally averaged differences given by equation  (\ref{eq:OLA_c_est}) and 
the corresponding equation for $\overline{\delta_x (G \rho)/(G \rho})$.  In practice that is not 
{entirely} straightforward for achieving the precision required in this endeavour.

We refer to the space spanned by the kernels $(K_{\covr,\rho}^{(n,l)},K_{\rho,\covr}^{(n,l)})$ as the kernel space; its orthogonal complement
is called the annihilator.  
Because 
relation (\ref{eq:annihilator})
can be interpreted as the vanishing of the inner product of the kernel
vectors
\begin{equation}
 \begin{pmatrix}
  K_{\covr,\rho}^{(n,l)}
\\
  K_{\rho,\covr}^{(n,l)}
 \end{pmatrix}
\label{eq:kernel_vector}
\end{equation}
and
the annihilator vector 
\begin{equation}
\ba 
 :=
\begin{pmatrix}
\displaystyle
 \frac{{\rd}\ln\covr}{{\rd}\ln r}
\\[3mm]
\displaystyle
 \frac{{\rd}\ln \rho}{{\rd}\ln r}
\end{pmatrix}\;
\label{eq:annihilator_comp}
\end{equation}
for all modes, we can say that 
the 
kernel vectors
given by
equations
(\ref{eq:av_c_kernels})
and
(\ref{eq:av_rho_kernels})
are orthogonal to
the annihilator vector $\ba$.
In fact, we easily find
from equation (\ref{eq:annihilator}) that
\begin{align}
&
\int
 \mathcal{K}^{(\covr)}_{\covr,\rho}
\,
 \frac{{\rd}\ln\covr}{{\rd}\ln r}
 \dx
+
\int
 \mathcal{K}^{(\covr)}_{\rho,\covr}
\,
 \frac{{\rd}\ln \rho}{{\rd}\ln r}
 \dx
\nonumber
\\
&=
\sum_{n,l} a_{n,l}
\left(
\int
 K_{\covr,\rho}^{(n,l)}
\,
 \frac{{\rd}\ln\covr}{{\rd}\ln r}
 \dx
+
\int
 K_{\rho,\covr}^{(n,l)}
\,
 \frac{{\rd}\ln \rho}{{\rd}\ln r}
 \dx
\right)
\nonumber\\&
=
0
\;.
\label{eq:avk_orth}
\end{align}
This means that 
inferences from OLA such as that
given by equation (\ref{eq:OLA_c_est}),
without the total mass constraint
(\ref{eq:mass_constraint}),
are never {influenced by} the annihilator {vector}
(\ref{eq:annihilator_comp})
of the target quantities
$\delta_x \covr/\covr$
and
$\delta_x (G\rho)/(G\rho)$.
We note that
at least one component of
the annihilator vector
(\ref{eq:annihilator_comp})
is
quite large, 
or
even
divergent,
at the stellar surface 
where pressure, density and temperature are all very small.
This is
characterized 
by the fact
that
the norm 
\begin{equation}
I_{\text{A}}(x_0)
:=
 \int_0^{x_0}
\left\{
\left(
\frac{{\rd}\ln \covr}{{\rd}\ln r}
\right)^2
+
\left(
\frac{{\rd}\ln \rho}{{\rd}\ln r}
\right)^2
\right\}
\dx
\;,
\label{eq:sq_norm_A}
\end{equation}
can be extremely large, or even divergent, 
as $x_0$ approaches its `surface' value, which we 
denote by $\xsurf$.
%
We are mindful to define
$\xsurf$
formally
as the vanishing point of the density distribution, 
recognizing that its value might be infinite if there were no distinct surface. 
If that be so, then $I_{\text{A}}(\xsurf)$ would also be formally infinite; however, 
its role in determining $\delta R/R$, as in equation (\ref{eq:dR_p1}) below, is 
via a non-divergent limiting process of the ratio of two individually divergent integrals.


We note that
there must exist
annihilator vectors other 
than the one given by
equation (\ref{eq:annihilator_comp})
that satisfy
orthogonality relations similar to 
equation
(\ref{eq:annihilator}).
For example,
if
the kernels of the eigenmodes included in the structure inversion 
are all negligibly small
in some region,
which typically
happens
outside
of the propagation cavities,
the structure in those regions
cannot be
probed by the eigenmodes.
Therefore in practice the structure difference in such regions
should be attributed to the annihilator.
There are also annihilator vectors
that are large within the propagation cavities, but if a wide variety of modes are included in the inversions, 
they are likely to be highly oscillatory
\citep[e.g.][]{wiggins1972RGSP10...251,Gough:1985aa}.

{Finally, we emphasize that, as evinced by 
equation (\ref{eq:perturbed_VP}), any component of 
the actual structure-difference vector 
$(\delta_x \psi/\psi, \delta_x (G\rho)/G\rho)$ 
that is orthogonal to all the kernel vectors 
$(K_{\covr,\rho}^{(n,l)},K_{\rho,\covr}^{(n,l)})$ 
makes no 
{contribution}
to the frequency data 
$\delta \nu _{n,l}/\nu _{n,l}$, and so  
cannot be inferred seismologically.  Therefore 
the  
{seismologically}
accessible element of the 
structure-difference vector must lie in the kernel 
space, and can be represented as a linear combination 
of the kernel vectors, 
as in equation (\ref{eq:s_kernel_space_component}).  
This forms the basis of 
some regularized least-squares data-fitting inversion 
procedures.  It is also explicit in OLA 
(equation(\ref{eq:OLA_c_est})).}
{%
In fact,
since
the localized averages
are totally insensitive
to the annihilator vectors
included in
the structure-difference vector,
the averages can be interpreted
as those of
only the element of
the difference vector in
the kernel space.
}

\section{Seismic radius inversion based on p-mode frequencies}
\label{sec:R_inversion}

\subsection{An inversion for the scale factor}
\label{subsec:OLA_inv_dR}
We first describe a method, based on a procedure 
analogous to the OLA 
inversion
for the structure variables, to infer
the radius difference defining the scale factor  $1+\delta R/R$.

By making
a linear combination of
equations
(\ref{eq:perturbed_VP})
and
(\ref{eq:mass_constraint}) with  coefficients $c_{n,l}$,  
we obtain
\begin{equation}
 \frac{\delta R}{R}
=
 \sum_{n,l} c_{n,l} \frac{\delta \nu_{n,l}}{\nu_{n,l}}
 - \mathcal{C}
-
\mathcal{S}
+ \frac{1}{3}\frac{\delta (GM)}{GM}
\;,
\label{eq:dR_R_expression2}
\end{equation}
in which
\begin{equation}
 \mathcal{C}
  :=
 \int \mathcal{K}^{(R)}_{\covr,\rho} \frac{\delta_x \covr}{\covr}\dx
+
\int
\mathcal{K}^{(R)}_{\rho,\covr}
\frac{\delta_x (G\rho)}{G\rho}\dx
\;,
\label{eq:C_def}
\end{equation}
and 
$\mathcal{S}$ is the contribution from the surface terms, given by
\begin{equation}
\mathcal{S} := \sum_{n,l} c_{n,l} S_{n,l}
\;.
\end{equation}
The symbols
$\mathcal{K}^{(R)}_{\covr,\rho}$
and
$\mathcal{K}^{(R)}_{\rho,\covr}$
in equation
(\ref{eq:C_def})
are defined by
\begin{align}
 \mathcal{K}^{(R)}_{\covr,\rho}
 &
 {:=}%
\sum_{n,l} c_{n,l} K_{\covr,\rho}^{(n,l)}
\label{eq:K_R_c_av}
\\
\noalign{\noindent and}
 \mathcal{K}^{(R)}_{\rho,\covr}
 &
 {:=}%
\sum_{n,l} c_{n,l} K_{\rho,\covr}^{(n,l)}
+\frac{4\pi R^3 x^2 \rho}{3 M}
\;,
\label{eq:K_R_rho_av}
\end{align}
{respectively.}
Note that the total-mass constraint is included here.
{This formulation is
formally similar to the one
for the mean-density inversion
by \citet{Reese:2012aa} in
asteroseismology.
The differences are that
the value of $GM$ is accurately known
for the Sun,
and that
we adopt $\covr$ and $G\rho$ as the target structure variables, while
{\citeauthor{Reese:2012aa}} 
adopt
$\rho$ and $\gamma_1$.}

The coefficients 
$c_{n,l}$
are determined as a {trade-off} between minimizing the magnitude of 
$\mathcal{C}$
and limiting the magnification of the frequency errors, 
simultaneously
%
 preventing
$\mathcal{S}$ from influencing the inferences.
%
This is accomplished by
minimizing with respect to $c_{n,l}$ the quantity
\begin{equation}
 \chi^2_R := \alpha_R \int ({\mathcal{K}^{(R)}_{\covr,\rho}})^2 \dx
        + \beta_R \int ({\mathcal{K}^{(R)}_{\rho,\covr}})^2 \dx
        + \gamma_R {\sigma}^2\;
\label{eq:chi2_R}
\end{equation}
under the condition that a representation of $\mathcal{S}$, which is defined below, 
vanishes;   
$\sigma$ is a formal error (also defined below, by equation
(\ref{eq:sigma_definition2})), 
and
$\alpha_R$, $\beta_R$ and $\gamma_R$ are
adjustable parameters.
An estimate of $\delta R/R$
is then given by
\begin{equation}
\left(
\frac{\delta R}{R}
\right)_{\text{ac}}
 {:=}
\sum_{n,l} c_{n,l} \frac{\delta\nu_{n,l}}{\nu_{n,l}}
 \pm \sigma \;,
\label{eq:dR_estimate2}
\end{equation}
where we have assumed that $\delta(GM)/(GM)=0$: 
{we are constructing a representation of the Sun with precisely the same value of $GM$ 
as that of our reference model, although we appreciate that that value may be in error by an amount 
of order $\sigma_{GM}$}. 
The formal error $\sigma$
is given by
\begin{equation}
 \sigma^2 := \sum_{n,l} \left(c_{n,l} \sigma_{n,l}\right)^2
+ \left(\frac{\sigma_{GM}}{3}\right)^2 \;,
\label{eq:sigma_definition2}
\end{equation}
in which 
$\sigma_{GM}$ and $\sigma_{n,l}$ denote
the relative observational standard errors
in the product $\GMsun$
(cf.~{Table}~\ref{tab:global_quantities}) and the 
frequencies $\nu_{n,l}$, respectively.  
We recall that $(\delta R/R)_{\text{ac}}$ defines a scaling based on the acoustic
structure of the star, and accordingly we have adopted the subscript `ac'.

In carrying out the radius inversions we take special care with the surface term 
by taking account of its leading $l$ dependence
\citep{Gough:1995aa,Di-Mauro:2002aa}.
The explicit expression {we adopt} is
\begin{equation}
S_{n,l} = \frac{1}{I_{n,l}}\left\{
 F_0(\nu_{n,l}) + \left(\frac{l + 1/2}{\nu_{n,l}}\right)^2 F_2(\nu_{n,l})\right\}\;,
\label{eq:surface_term}
\end{equation}
where $I_{n,l}$ is
the mode inertia normalized by that
of the radial mode with {almost} the same frequency
\citep[cf.][]{Christensen-Dalsgaard:1991aa},
and the functions
$F_0$ and $F_2$ are arbitrary, and depend only on frequency.
Both $F_0$ and $F_2$ are expanded as series of {$n_0$ and $n_2$} Legendre
polynomials{, respectively,}
whose arguments
are normalized so that
the whole range of the frequencies that are included in the representation 
corresponds to the interval $[-1, 1]$.
To ensure that $\mathcal{S}$ vanishes, we adopt  
Lagrange's method of 
{undetermined multipliers} 
to obtain the coefficients in the Legendre expansions 
during the  minimization of $\chi^2_R$. 
The resulting values of $(\delta R/R)_{\text{ac}}$, for several values 
of 
{
$n_0$ and $n_2$
}, 
are listed {below} in {Table} \ref{tab:R_surface_term_effect}.

{
Although
our formulation
does not {explicitly} distinguish p modes from f modes 
(nor, even, g modes), in practice 
it is not a good idea
to mix both types of modes in the
structure and radius inversion, 
because
the functional form of
the surface term
for p modes 
is likely to
be qualitatively different from that for f modes, owing to 
their different responses to near-surface perturbations 
\citep[cf.][]{Gough:1993aa,Di-Mauro:2002aa}, as we point out in {Section} \ref{sec:real_R_inv}.
}%

\subsection{Interpretation of the inverted 
\texorpdfstring{$\delta R/R$}{δR/R}}
\label{subsec:interpretation_of_dRR}

Although a formal procedure to obtain  
$(\delta R/R)_{\text{ac}}$ given by equation (\ref{eq:dR_estimate2}) 
has been developed in {Section}
\ref{subsec:OLA_inv_dR},
we still need to consider
how to interpret the result.
To this end 
we examine
the requirement of the inversion procedure in {Section} \ref{subsec:OLA_inv_dR}
that the contribution
from $\mathcal{C}$
to $\delta R/R$ 
{be rendered} negligible
{(}cf.~equation (\ref{eq:dR_R_expression2}){)}.
The outline of the discussion is as follows:
though it might appear at first sight
that this requirement can be satisfied easily
if the frequencies of
a sufficient number of eigenmodes with different characters
are available,
it is actually impossible to satisfy
it
if the annihilator vector
$\ba$ {seriously} `contaminates'
the relative structure difference,
$\left( \delta_x \covr/\covr,\,
 \delta_x \left(G\rho\right)/\left(G\rho\right) \right)$;
this contamination can be removed
only if the definition of $x$, or equivalently $\delta R/R$, is
adjusted appropriately,
and we claim that
only in this case does the inversion procedure provide a meaningful
answer;
we interpret
the {outcome} of the inversion given by equation (\ref{eq:dR_estimate2})
as {that} value of $\delta R/R$ that makes
the relative structure difference {almost} free from
the annihilator vector $\ba$, 
{as we now discuss.}

Perusal of 
equation (\ref{eq:dR_R_expression2}) for $\delta R/R$ 
reveals that the right-hand side {itself} 
depends on $\delta R$, via the difference operator $\delta_x$ that appears in $\mathcal{C}$.  The transformation between  $\delta_x$ and $\delta_r$ 
is given explicitly by 
\begin{equation}
\begin{pmatrix}
\displaystyle
 \frac{\delta_r c}{c}
\\[3mm]
\displaystyle
 \frac{\delta_r (G \rho)}{G \rho}
\end{pmatrix}
 = 
\begin{pmatrix}
\displaystyle
 \frac{\delta_r \covr}{\covr}
\\[3mm]
\displaystyle
 \frac{\delta_r (G \rho)}{G \rho}
\end{pmatrix}
 = 
\begin{pmatrix}
\displaystyle
 \frac{\delta_x \covr}{\covr}
\\[3mm]
\displaystyle
 \frac{\delta_x (G \rho)}{G \rho}
\end{pmatrix}
-
\frac{\delta R}{R}
\ba.
\label{eq:drdx_dif2}
\end{equation}
It is important to realize
that the decomposition
on the right-hand side of
equation (\ref{eq:drdx_dif2})
is 
in a sense
arbitrary, 
as we discussed in the introduction.    
To render it {determinate} requires another constraint,
 which we are free to choose at will.
We pay attention to
the main assumption upon which equation (\ref{eq:dR_estimate2}) relies: 
that with enough modes of sufficiently diverse variety  
the  contribution
from $\mathcal{C}$ can be made negligible; the coefficients $c_{n,l}$ have been 
determined in such a way as to make $\mathcal{S}$ vanish, and the uncertainty in $GM$ 
is much smaller than that of the other quantities influencing our analysis, so that, 
aside from frequency-measurement errors, the  
term  $\mathcal{C}$ contaminating the approximation (\ref{eq:dR_estimate2}) 
to equation (\ref{eq:dR_R_expression2})
is all that 
remains to degrade the estimate.  
An essential point is that, 
because the magnitude of  
 the annihilator vector $\ba$ given by
equation (\ref{eq:annihilator_comp})  is large near the stellar surface,
if it is included in
the fractional structure difference
at fixed $x$,  expressed by
the second term on the right-hand side of
equation (\ref{eq:drdx_dif2}), 
its contribution to $\mathcal{C}$
is not negligible,
however small
we can make $\mathcal{K}^{(R)}_{\covr,\rho}$ and
$\mathcal{K}^{(R)}_{\rho,\covr}$ 
by adjusting the coefficients $c_{n,l}$.
This can be understood explicitly from the 
following expression for the projection of the annihilating vector function $\ba$ 
onto the kernels of $\mathcal{C}$:
\begin{align}
& \int \mathcal{K}^{(R)}_{\covr,\rho} \frac{{\rd}\ln \covr}{{\rd}\ln r}\dx
+
\int
\mathcal{K}^{(R)}_{\rho,\covr}
\frac{\rd\ln\rho}{\rd\ln r}\dx
\nonumber\\
& =
\int
\frac{4\pi R^3 x^2 \rho}{3 M}
 \frac{\rd\ln\rho}{\rd\ln r}\dx
 = -1
\;, 
\label{eq:a_dot_b}
\end{align}
which can be derived from equations
(\ref{eq:K_R_c_av}), (\ref{eq:K_R_rho_av}),
(\ref{eq:annihilator}),
and
(\ref{eq:mass_conservation}),
assuming that density vanishes, or is at least negligible, at the surface. 
Evidently, that projection is not small. 
In other words,
the assumption of negligible contribution from $\mathcal{C}$
cannot be justified
without adjusting the decomposition given by equation
(\ref{eq:drdx_dif2}).

To obtain an appropriate adjustment
we explicitly distinguish
the seismologically accessible 
{element}
from
the annihilator vector thus:
\begin{equation}
 \begin{pmatrix}
\displaystyle  
 \frac{\delta_r \covr}{\covr}\\[3mm]
\displaystyle  
 \frac{\delta_r \left(G\rho\right)}{G\rho}
 \end{pmatrix} 
 =
\begin{pmatrix}
\displaystyle  
 \frac{\delta_r \covr}{\covr}\\[3mm]
\displaystyle  
 \frac{\delta_r \left(G\rho\right)}{G\rho}
\end{pmatrix}_{\text{ker}}
+
\begin{pmatrix}
\displaystyle  
 \frac{\delta_r \covr}{\covr}\\[3mm]
\displaystyle  
 \frac{\delta_r \left(G\rho\right)}{G\rho}
\end{pmatrix}_{\text{a}}
\;,
\label{151056_17Oct18}
\end{equation}
where
\begin{equation}
\begin{pmatrix}
\displaystyle  
 \frac{\delta_r \covr}{\covr}\\[3mm]
\displaystyle  
 \frac{\delta_r \left(G\rho\right)}{G\rho}
\end{pmatrix}_{\text{ker}}
:=
\sum_{n,l}
k_{n,l}
\begin{pmatrix}
 K_{\covr,\rho}^{\left(n,l\right)}\\
 K_{\rho,\covr}^{\left(n,l\right)}
\end{pmatrix}
\label{eq:s_kernel_space_component}
\end{equation}
is the seismologically estimated 
{accessible} 
{element} 
of the actual
{structure difference}
$\left(
\delta_r \covr/\covr,\,\delta_r\left(G\rho\right)/\left(G\rho\right)
\right)$
that lies in the kernel space,
from which we wish to achieve a reliable estimate of
$\mathcal{C}$.
{Equation (\ref{eq:s_kernel_space_component}) is sometimes 
called a spectral expansion
{\citep[e.g.~Section 41.1.1 of][]{UOASS1989}}.}
Note that the second term
on the right-hand side of equation (\ref{151056_17Oct18})
can generally
include
not only a term proportional to $\mathbf{a}$, 
but also other vectors in the annihilator.
{
We note {also} that
{$({\delta_r \psi}/{\psi},{\delta_r \left(G\rho\right)}/{G\rho})_{\text{ker}}$}
corresponds to the projection of the structure difference
onto the kernel space.
The reason why it is seismologically accessible 
is that the right-hand side of
equation (\ref{eq:perturbed_VP}),
without the surface term $S_{n,l}$,
can be interpreted as a nontrivial inner product
of the structure difference with the kernel vectors.
}
{On} substituting
equations (\ref{eq:drdx_dif2}) and (\ref{151056_17Oct18})
into equation (\ref{eq:C_def}),
we obtain
with the help of equation (\ref{eq:a_dot_b})
\begin{equation}
 \mathcal{C}
  =
  \mathcal{C}_{\text{ker}}
  -
  \frac{\delta R}{R}
  +
  \int
  \frac{4\pi R^3 x^2\rho}{3 M}
    \left(
   \frac{\delta_r \left(G\rho\right)}{G\rho}
   \right)_{\text{a}}
  \dx
  \;,
\label{152647_17Oct18}  
\end{equation}
in which
\begin{equation}
 \mathcal{C}_{\text{ker}}
  :=
  \int
  \mathcal{K}_{\covr,\rho}^{(R)}
  \left(
   \frac{\delta_r \covr}{\covr}
  \right)_{\text{ker}}
  \dx
  +
  \int
  \mathcal{K}_{\rho,\covr}^{(R)}
  \left(
   \frac{\delta_r\left(G\rho\right)}{G\rho}
  \right)_{\text{ker}}
  \dx
\;.  
\end{equation}
Since $\mathcal{C}_{\text{ker}}$
depends on
only the seismologically accessible 
{element},
we may adopt
as the fundamental assumption
of the inversion procedure in {Section} \ref{subsec:OLA_inv_dR}
that
$\mathcal{C}_{\text{ker}}$ can be made
negligibly small
by adjusting the coefficients $c_{n,l}$ appropriately
if
we have a sufficient number of
eigenmode kernels with different characters.
Since the estimate for $\delta R/R$
given by equation (\ref{eq:dR_estimate2})
is meaningful
only when $\mathcal{C}$ is negligible as a whole,
we interpret, 
based on
equation (\ref{152647_17Oct18}), 
that
it is the estimate
in the case where
$\delta R/R$ is set to
\begin{equation}
\left(
 \frac{\delta R}{R}
 \right)_{\text{interpret}}
  =
  \int
  \frac{4\pi R^3 x^2\rho}{3 M}
  \left(
   \frac{\delta_r \left(G\rho\right)}{G\rho}
   \right)_{\text{a}}
  \dx
  \;.
\label{153159_17Oct18}  
\end{equation}
Seemingly paradoxically {at first},
equation
(\ref{153159_17Oct18})
appears to imply that the relative difference in the scale factor is
determined by the 
element
of the density-profile difference that is not accessible to the eigenfrequencies.  {Of course, that} is fully consistent with the isospectral nature of the problem that is discussed in {Sections} \ref{subsec:annihilator_relation} and 
 \ref{subsec:radius_determination}.

Finally, we note that the minimization of the influence of the contaminating integral
$\mathcal{C}$
{in} expression (\ref{eq:dR_R_expression2}) for
$\delta R/R$ answers question \ref{it:Q2}
at the end of {Section} \ref{subsec:annihilator_relation}.

\subsection{Mode-set independent interpretation}
\label{015032_29Oct18}

Because the annihilator 
necessarily
depends
on the mode set available, so does 
the interpretation given by
equation
(\ref{153159_17Oct18}).
To obtain
an expression for $\delta R/R$ that is 
{only weakly} dependent of the mode set,
we consider 
there to be such a 
large variety of modes {available} 
that
the annihilator space
can be assumed to be composed of
only the vector $\ba$ defined by
equation (\ref{eq:annihilator_comp})  
-- together, of course, with the highly oscillatory functions which we 
accept cannot be resolved, and which accordingly we ignore.
This assumption permits replacing 
equation
(\ref{151056_17Oct18})
by 
\begin{equation}
\begin{pmatrix}
\displaystyle
 \frac{\delta_r c}{c}
\\[3mm]
\displaystyle
 \frac{\delta_r (G \rho)}{G \rho}
\end{pmatrix}
 = 
\begin{pmatrix}
\displaystyle
 \frac{\delta_r c}{c}
\\[3mm]
\displaystyle
 \frac{\delta_r (G \rho)}{G \rho}
\end{pmatrix}_{\text{ker}}
-
\frac{\delta R}{R}
\ba
\label{eq:dstruc_limit}
\end{equation}
{(cf.~equation (\ref{eq:drdx_dif2})).}   
{Recognizing} that both components of $\ba$
are {typically} quite large
near the surface,
and {that}  
the frequency kernels, which are the constituents
of the spectral expansion,
are very small, 
we neglect the first term on the right-hand side of 
equation (\ref{eq:dstruc_limit}) and   
are led to the estimates
\begin{equation}
 \left(
  \frac{\delta R}{R}
  \right)_{c}
 {:=}
\lim_{x_0\rightarrow\xsurf}  
\frac{\displaystyle
\frac{\delta_r c}{c}(x_0)}
{\displaystyle
 -\frac{{\rd}\ln \covr}{{\rd}\ln r}(x_0)}
\label{eq:dRR_c_inf} 
\end{equation}
and
\begin{equation}
 \left(
  \frac{\delta R}{R}
  \right)_{\rho}
 {:=}
\lim_{x_0\rightarrow\xsurf}  
\frac{\displaystyle
\frac{\delta_r (G\rho)}{G\rho}(x_0)}
{\displaystyle
 -\frac{{\rd}\ln\rho}{{\rd}\ln r}(x_0)}
 \;.
\label{eq:dRR_rho_inf} 
\end{equation}
We may expect 
equation (\ref{eq:dRR_rho_inf}) {to} converge
faster
than
equation (\ref{eq:dRR_c_inf})
because $\left|{\rd}\ln\rho/{\rd}\ln r\right|$ is much larger
than $\left|{\rd}\ln\covr/{\rd}\ln r\right|$
in realistic stellar structures.  
In fact, $\left|{\rd}\ln\covr/{\rd}\ln r\right|$ can {actually} be 
{quite} small 
in the vicinity of the temperature minimum, 
in which case our argument {formally} breaks down {because} equation (\ref{eq:dRR_c_inf}) 
is not well satisfied.

\begin{figure}
\centering\includegraphics[width=\columnwidth]{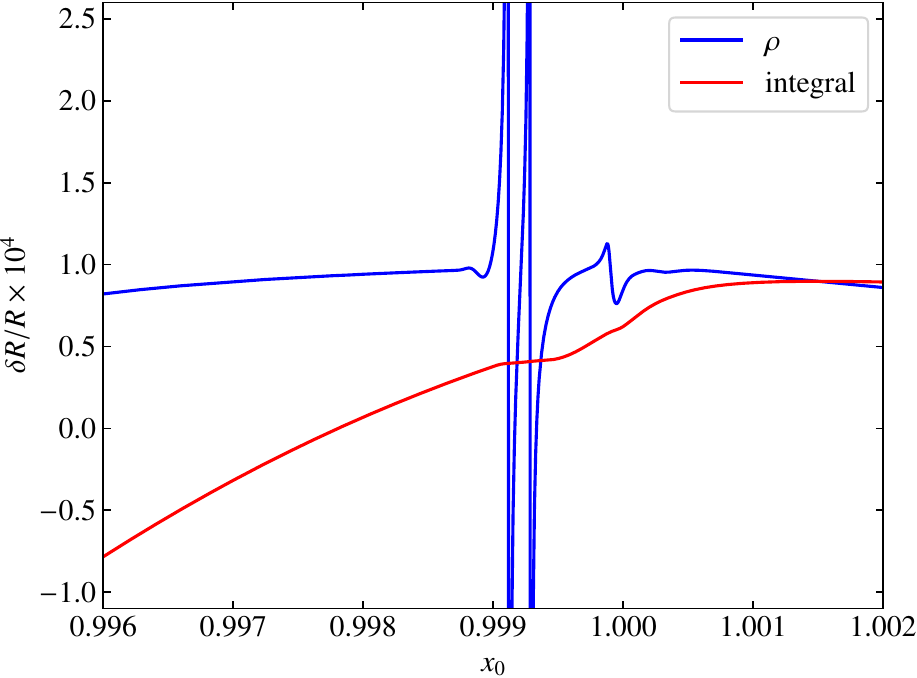}
\caption{
Estimates of $\delta R/R$ given by equation 
(\ref{eq:dRR_rho_inf}) 
{(blue)}
and 
equation (\ref{eq:dR_p1}) 
{(red)}
relating a target theoretical 
solar model to a reference model, {model} S of  
\citet{Christensen-Dalsgaard:1996aa}, plotted against 
{the value} 
{$x_0$}
{of $r/\Rph$}
in the 
outer layers. 
{Here, 
$\Rph$ means
the photospheric radius.}
The target model is essentially {model} 1 of 
\citet{HGseconddiff2007MNRAS}, 
scaled to a photospheric radius 
{$1.0001\,
\Rph$}. Additional information 
about that model is provided in {Section} 
{\ref{subsec:numerical_tests_for_R}.}
}
\label{fig:figRformulae}
\end{figure}

Note that by taking the inner product of  
equation (\ref{eq:dstruc_limit}) 
with $\ba$, we obtain 
\begin{align}
 &
\left(
 \frac{\delta R}{R}
\right)_{\text{integral}}
 {:=}
\nonumber\\& 
\lim_{x_0\rightarrow\xsurf}  
\frac{-1}{I_{\text{A}}(x_0)}
\int_0^{x_0}
\left(
 \frac{\delta_r c}{c}\frac{{\rd}\ln\covr}{{\rd}\ln r}
+
 \frac{\delta_r (G \rho)}{G \rho}\frac{{\rd}\ln\rho}{{\rd}\ln r}
\right)
\dx
\;,
\label{eq:dR_p1}
\end{align}
in which $I_{\text{A}}(x_0)$
is defined by equation (\ref{eq:sq_norm_A}).
%
One might expect this expression to be
 more robust because it is somewhat less susceptible to the details of the relatively small-scale variation of the structure of the star near its surface  (see Section \ref{subsec:numerical_tests_for_R_form}).
That is indeed the case, as is illustrated in 
Fig. \ref{fig:figRformulae}, in which the right-hand sides of equations 
(\ref{eq:dRR_rho_inf}) and (\ref{eq:dR_p1}), obtained from 
the difference between two 
theoretical solar models, are plotted against 
{$x_0$} 
in the outer layers.  We have not included 
{a corresponding estimate from equation (\ref{eq:dRR_c_inf}), but merely report that 
it oscillates more wildly than the estimate from equation (\ref{eq:dRR_rho_inf}).  We emphasize 
that the resulting total radius is, in principle, independent of the reference model from which it 
was obtained.} 

\subsection{%
{Relation to the acoustic radius}
}
\label{subsec:rel_to_acr}

{
We finally consider how to
interpret
$\left(\delta R/R\right)_{\text{ac}}$,
given by
equation (\ref{eq:dR_estimate2}),
in terms of the seismic radii,
which are introduced in
{Section} \ref{sec:seismic_radii}.
}
{
As is 
{{illustrated} by \citet{Takata:2003ab},} 
the density profile
near the surface,
where the kernels start to decay outwards in response to 
acoustic reflection,
is crucial in determining 
{the}
{p-scaled}
radius difference $\delta R/R$; 
that is true also
for
the 
{f-scaled}
radius
\citep{Schou:1997aa,Dziembowski:2001aa}.
%
%
%
%
It is therefore clear
that
the density profile in the vicinity of 
the upper turning points is
the most important characteristic for  determining the 
{p-scaled}
radius.
In fact,
it tells us that
the difference in the 
{p-scaled}
radius
can be interpreted
as the
homologous difference
in the density profile
around the upper turning points.
This means that,
if the density $\rho$ itself becomes
much smaller than the density difference $\delta_r \rho$
near the surface,
the inversion procedure
(without the total mass constraint)
attributes
the large relative difference $\delta_r \rho/\rho$
to the homologous difference
so that
the scaled difference
$\delta_x (G\rho)/(G\rho)$
is prevented from
being too large near the surface.
It seems as if
the inversion procedure
adjusts
the scale factor $R$
to render 
the assumption of the linearization to be as near to being 
valid as possible.}

{%
In order to understand further 
the relation between
the density and the radius difference,
we summarize
the density profile
near the surface of the Sun
based on {model} S.
The atmosphere 
around the temperature minimum,
which is located $\sim 500~\text{km}$
above the photosphere,
can be well approximated by
an isothermal structure,
which implies
the density decreases exponentially outwards
with a constant scale height.
The scale height gradually
gets larger {with diminishing height above} the photosphere,
where it is approximately equal to
${180}~\text{km}$
(${2.6}\times 10^{-4}$ in fractional radius).
{Beneath} the photosphere,
it first increases very rapidly
until it reaches its local maximum at {a depth of} 
$\sim 80~\text{km}$ below the photosphere,
and then decreases steeply
until it takes its local minimum at {a depth of} 
$\sim {250}~\text{km}$,
after which it turns to increase
inwards mildly.
Note that 
{the seismic radius $\Rc$,
which was described in {Section} \ref{sec:seismic_radii},
is located very close to
the local maximum of the density scale height.}
Because physical processes
in the atmosphere can be understood
relatively well,
we may assume that
the {density-profile difference} 
between the Sun and the reference model
is small 
between the photosphere
and the temperature minimum,
except 
for {a} possible displacement 
due to
the position difference in the photosphere.
Since $\Rph$ is quite close to $\Rc$,
the difference 
in the atmosphere
can largely be removed by
stretching (or contracting)
the radial coordinate of
the reference model {to make} 
$\Rc$ 
{consistent with} the Sun. 
This essentially corresponds to
{eliminating}
the annihilator component%
{,
which is represented by
the second term on the right-hand side of
equation (\ref{eq:dstruc_limit}),
}
from the density difference.
We thus identify} 
\begin{equation}
\left(\frac{\delta R}{R}\right)_{\text{ac}} = \frac{\delta\Rc}{\Rcr}
\label{eq:dRac_dRc}
\end{equation}
{
(cf.~equation (\ref{eq:scaling_assumption})).
Note that
there could
remain {a} small difference
in the density structure above $\Rc$,
even after {the adjustment.}
This
may originate from
uncertainties in the
description of the superadiabatic 
convective layers,
which {contain} $\Rc$.
}

\section{
{
Structure Inversions
with radius difference}}
\label{sec:revision}

Because in conventional structure inversions 
the difference between
the 
{photospheric}
radius {of}
the Sun 
and that of the reference model
is {usually} ignored  \citep[see][]{Richard:1998aa}, 
it behoves us now to offer a  
modification. 
We concentrate particularly on inversions for
sound speed 
 and density.
{We expand,}
in {Sections} \ref{subsec:mod_c_inv} and \ref{subsec:mod_rho_inv},
a method 
to obtain
inferences
 about
$\delta_x c/c$ and
$\delta_x (G\rho)/(G\rho)$
which depend on $\delta R/R$
only implicitly
through the definition of $x=r/R$; 
{then} we can  determine
the radius difference $\delta R/R$ independently 
by the method described in {Section} \ref{sec:R_inversion}.


\subsection{Formulation of the sound-speed inversion}
\label{subsec:mod_c_inv}

The basic equation of the analysis
can be obtained as a linear combination of 
equations (\ref{eq:perturbed_VP})
and (\ref{eq:mass_constraint}), 
using {equations (\ref{eq:covr_def}) and} (\ref{eq:dxr_dR}), yielding
\begin{align}
\sum_{n,l} \invce_{n,l} \frac{\delta \nu_{n,l}}{\nu_{n,l}}
&=
\int \mathcal{K}^{(c)}_{c,\rho} \frac{\delta_x c}{c} \dx
+
\int
\mathcal{K}^{(c)}_{\rho,c}
\frac{\delta_x (G\rho)}{G\rho} \dx
\nonumber\\ &\phantom{=}\mbox{}
+
\left(
1 - \int \mathcal{K}^{(c)}_{c,\rho}\dx
\right)
\frac{\delta R}{R}
-
\frac{1}{3}
\frac{\delta (GM)}{GM}
+
\mathcal{S}
\;,
\label{eq:c_inv_basic_eq}
\end{align}
in which
$\mathcal{K}^{(c)}_{c,\rho}$ and
$\mathcal{K}^{(c)}_{\rho,c}$ are
{defined by}
\begin{equation}
\begin{pmatrix}
 \mathcal{K}^{(c)}_{c,\rho}
 \\
 \mathcal{K}^{(c)}_{\rho,c}
\end{pmatrix}
:=
 \sum_{n,l} \invce_{n,l} 
\begin{pmatrix}
K_{c,\rho}^{(n,l)}
\\
K_{\rho,c}^{(n,l)}
\end{pmatrix}
+
\begin{pmatrix}
0
\\
\frac{4\pi R^3 x^2 \rho}{3M}
\end{pmatrix}
\, .
\label{eq:av_c_kernels_rev}
\end{equation}
{%
Here, $\invce_{n,l}$ are {potential} inversion coefficients.
}
Within the framework of
the SOLA inversion
\citep{Pijpers:1992aa}
an inference about the sound speed 
at the fractional radius $\xt$ can be made
by minimizing the quantity 
\begin{equation}
 \chi^2_c {:=}
 \int
\left[
\mathcal{K}^{(c)}_{c,\rho} - T(x;\xt)
\right]^2
  \dx
        + \alpha_c \int \left({
        \mathcal{K}^{(c)}_{\rho,c}
        }\right)^2 \dx
        + \beta_c \sigma^2
\label{eq:c_chi2}
\end{equation}
under the 
unimodular
normalization condition 
\begin{equation}
 \int \mathcal{K}^{(c)}_{c,\rho} \dx = 1
\label{eq:c_normalization}
\end{equation}
together with the vanishing of
the surface term.
Here the 
target {kernel} 
$T(x;\xt)$ is 
usually {taken to be}  
a Gaussian function centred at $x=\xt$
with an adjustable width
chosen
so as
to limit
undue 
magnification of the data errors.
Positive {constants}  
$\alpha_c$ and $\beta_c$ are adjustable parameters, 
and
the formal error $\sigma$ is still given by
equation (\ref{eq:sigma_definition2})
(with $c_{n,l}$ replaced by $\invce_{n,l}$).

{In deriving equation (\ref{eq:c_inv_basic_eq})
we have transformed 
the dependent variable $\delta_x \psi$ 
in equation (\ref{eq:perturbed_VP})
to $\delta_x c$ according to 
$\delta_x\covr/\covr = \delta_x c/c - \delta R/R$, 
partly because $c$ is of greater interest to 
astrophysicists, and, interestingly, because}
the influence of $\delta R/R$ is {explicitly} removed 
from equation (\ref{eq:c_inv_basic_eq})
by condition (\ref{eq:c_normalization}), 
{which is adopted also, both here and in subsection \ref{subsec:annihilator}, 
as a convenient normalization of the localized averaging kernel.}

What we obtain by minimizing $\chi_c^2$ are
the inversion coefficients $\invce_{n,l}$,
from which
the difference in the sound speed 
in the vicinity of
$\xt$ is estimated by 
\begin{equation}
\left[
 \overline{\frac{\delta_x c}{c}}
\right]_{\text{OLA}}
{:=}
 \sum_{n,l} \invce_{n,l} \frac{\delta\nu_{n,l}}{\nu_{n,l}}
+ \frac{1}{3}\, \frac{\delta(GM)}{GM}
\label{eq:c_estimate}
\end{equation}
with the formal error $\sigma$.
The averaged sound-speed difference is to be regarded as 
a function of $x$.
Since
the product $GM$ of the Sun is measured very accurately 
(see {Table} \ref{tab:global_quantities}),
we can 
neglect the second term
on the right-hand side of equation (\ref{eq:c_estimate}).
Correspondingly,
the relative error 
$\sigma_{GM}$
contributes little to the formal error $\sigma$
in equation (\ref{eq:sigma_definition2}).
In fact, expressions
(\ref{eq:c_chi2})
and
(\ref{eq:c_normalization})
do not look new at all
because 
both of them are also found in conventional inversions.
The only difference is that
the coefficient of the total mass constraint
(\ref{eq:mass_constraint})
is {now} fixed at $1/3$ 
in the {current} formulation
when
we make
the linear combination
(\ref{eq:c_inv_basic_eq}),
{whereas it is} 
determined
by minimizing a quantity like
$\chi_c^2$ in equation (\ref{eq:c_chi2})
in conventional inversions.
We should stress that 
by repeating the argument presented in {Section} \ref{subsec:interpretation_of_dRR} it follows that 
the interpretation
of $\delta R/R$ here is the same as that given by equation
(\ref{153159_17Oct18}); and it applies also
for the density inversion {addressed} in the next section.


\subsection{Formulation of the density inversion}
\label{subsec:mod_rho_inv}

Unlike in the 
{sound-speed}
inversion,
we should not include 
in the density inversion
the total mass constraint (\ref{eq:mass_constraint}),
which contains
a term proportional to $\delta R/R$; 
we need only equation (\ref{eq:perturbed_VP})
to get inferences of $\delta_x (G\rho)/(G\rho)$
without the explicit effect of $\delta R/R$.
Actually,
the procedure 
is simply the same as
 the conventional one without the total mass constraint.
The most important difference between the two procedures
is in the interpretation of the results: 
namely,
what is regarded as $\delta_r \rho/\rho$
in the conventional method
should be recognized as $\delta_x (G\rho)/(G\rho)$
in the new method.
Because the gravitational constant $G$
is always 
multiplied by
$\rho$, or, equivalently $M$, 
the uncertainty in $G$ itself cannot
affect the inversion process.
Therefore one
cannot {determine} by helioseismology alone whether or not 
$\delta G$ is zero.


\section{Numerical tests}
\label{sec:numerical_tests}

In this section,
we
test,
based on theoretical models,
the formulae for the radius difference proposed in {Section}
\ref{015032_29Oct18}, 
and the inversion procedures for 
the radius, sound speed and density
that are
developed in 
{Sections}
\ref{subsec:OLA_inv_dR},
\ref{subsec:mod_c_inv}
and
\ref{subsec:mod_rho_inv}, respectively.

\subsection{{Reference and target models}}
\label{subsec:numerical_tests_for_R}
 
{
We use two theoretical solar models in this section,
{model} S of \citet{Christensen-Dalsgaard:1996aa}
and
{model} 1 of \citet{HGseconddiff2007MNRAS},
which are adopted as reference and target models,
respectively.
}
Model 1 has an age of $4.15\,{\text{Gy}}$ and a heavy-element abundance 
$Z = 0.0200$; no gravitational settling was incorporated in its 
construction.  
By comparison, the age of {model} S is $4.60\,{\text{Gy}}$, the  
initial heavy-element abundance is $Z_0 = 0.01963$, and it has suffered 
gravitational settling.  
The models have the same photospheric radius: 
$\Rph=695.99$~Mm, 
consistent with
the value of \cite{Allen:1973aa} (see {Table} \ref{tab:global_quantities}), 
and the surface luminosity
($3.846\times 10^{33}$ erg $\text{s}^{-1}$);
they were constructed with different opacity tables.  
Both models 
were extended by smoothly adding isothermal atmospheres out to a 
radius of $1.002\,{\Rph}$.
The radius variable of the target model was then {multiplied} by
a factor 1.0001
{
in {Section} \ref{subsec:numerical_tests_for_R_form}
and
$0.9999$
in {Sections}
\ref{subsec:numerical_tests_for_R_inv}
and
\ref{subsec:numerical_tests_for_sturc_inv},} 
and its density and 
sound speed were scaled homologously. 

\subsection{
{
Expressions for the radius difference
}
}
\label{subsec:numerical_tests_for_R_form}

{
We remark here simply that evidence for the accuracy of 
the two explanatory
formulae (\ref{eq:dRR_rho_inf}) and (\ref{eq:dR_p1}) is presented in 
Fig. \ref{fig:figRformulae}.  
}
{
Both expressions
(\ref{eq:dRR_rho_inf}) and  (\ref{eq:dR_p1})
are functions that flatten with increasing height
in the atmosphere;}
the integral 
expression (\ref{eq:dR_p1}) tends to a constant, although 
$-(\delta_r (G \rho)/{G \rho})/({\rd}\ln\rho/{\rd}\ln r)$ 
{declines slowly}.
The relative difference between the photospheric radii of the two 
models is $10^{-4}$,
whereas the values obtained by the two formulae are both about
 $0.9 \times 10^{-4}$.
The $10^{-5}$ discrepancy, which arises at least in part from the 
non-homologous difference between the two models above their turning points, 
provides an indication of the accuracy of these formulae.

\subsection{Radius inversion}
\label{subsec:numerical_tests_for_R_inv}
We have performed a 
test calculation to assess
the accuracy of
the radius inversion method formulated in
{Section} \ref{subsec:OLA_inv_dR}.
To ensure that the computed oscillation eigenfrequencies faithfully represent 
the equilibrium models, 
slight adjustments were made to those models by recomputing hydrostatic 
balance to high
precision (to order $10^{-6})$, retaining the variation of density and buoyancy 
frequency as the defining properties.
The relative differences in the
sound speed and density 
between the adjusted and original models
are of the order of $10^{-5}$ or less
for $0.05 \le r/\Rph \le 0.95$,
where $\Rph$ means photospheric radius,
while the central values of the sound speed and density
are lower in the adjusted models by about 
$0.05$~{per cent} and $0.1$~{per cent}, respectively.
{In addition,
these evolutionary models are artificially extended
by $0.5$~{per cent} in radius
to the higher layers in the isothermal atmosphere
in order to ensure that the mode kernels have
negligibly small amplitude {at}  the outermost mesh point.
}
The structure of {model} 1
was shrunk homologously 
in the radial direction
by $0.01$~{per cent},
and
increased in density and sound speed at each value of 
the mass coordinate
by $0.03$~{per cent} and   {$0.005$}~{per cent},
respectively; {that implies a very slight alteration to
the equation of state.}
Since the  target model so constructed has the same mass as the
original model, but a smaller photospheric radius, by $0.01$~{per cent}, 
each mode frequency {is} augmented by $0.015$~{per cent}.
We point out that the structure difference between
the target and reference models is not homologous.

The difference between 
{the radii of} the two models was inferred
from the difference in their adiabatic eigenfrequencies,  
which were computed by
the program described by
\citet{Takata:2004ab}.
The mode set adopted 
was that of the MDI 360-day data
\citep{Schou:1999aa}, 
excluding the f modes.
We used in total 2008 eigenmodes 
with degrees ranging  from $l=0$ to $189$,
and having frequencies between $900$ and $4600$ mHz.
The frequency error of each mode
of the target model
was assumed
to be the same as
that of the corresponding mode
in the MDI data.  
The surface term was ignored in this test.
The inferred value of
${(}\delta R/R{)_{\text{ac}}}$ 
is $\left(-1.2 \pm 0.4 \right) \times 10^{-4}$
for a set of the parameters 
$\left(\alpha_R,\,\beta_R,\,\gamma_R\right)
= \left(1,\,300,\,2\times 10^{5}\right)$.
Because the frequencies used  are
free from observational errors,
the uncertainties in the inference (of $4\times 10^{-5}$)
should be regarded as being only formal.
On the other hand,
the systematic error,
which originates from the neglected nonlinear term
{of} the order of $(\delta\nu_{n,l}/\nu_{n,l})^2$
in equation (\ref{eq:perturbed_VP}),
could be estimated to be {of} the order $10^{-5}$.
{%
In order to examine the stability of the radius inversion,
we change the maximum spherical degree, $l_{\max}$, of the mode set, but keep the same values of 
$\alpha_R$, $\beta_R$ and $\gamma_R$.
Table~\ref{tab:rinv_lmax_var} shows
the results for $l_{\max}=120$, $140$, $160$ and $189$.
Although the results decrease 
{slightly} 
as $l_{\max}$ increases
from $l_{\max}=120$ to $l_{\max}=189$,
they are all consistent with each other within 
$2\times 10^{-5}$.
}
From 
{these test inversions},
we conclude that
the inferred relative difference
in the 
{p-scaled}
radius
is consistent with
that in the photospheric radius of $-1 \times 10^{-4}$
within the systematic error.
%

\begin{table}
\caption{
{
{Test radius}
inversions
{based on} 
{the procedure outlined in Section 
\ref{subsec:OLA_inv_dR} using the}
{two theoretical models}  {of} {{Section} \ref{subsec:numerical_tests_for_R}}, 
{the target having been shrunk homologously 
by $0.01$~{per cent}}, 
for various values of the maximum spherical degree,
$l_{\max}$.
The number of modes included in the data sets
is indicated by $N_{\text{mode}}$.
Note that only p modes are used.
}
}
\label{tab:rinv_lmax_var}
\centering
\begin{tabular}{ccc}
\hline
$l_{\max}$ & $N_{\text{mode}}$ & $(\delta R/R)_{{\text{ac}}} \times 10^4$\\
\hline
$120$ & $1719$ & $-1.0 \pm 0.5$
\\
$140$ & $1847$ & $-1.1 \pm 0.4$
\\
$160$ & $1945$ & $-1.2 \pm 0.4$
\\
$189$ & $2008$ & $-1.2 \pm 0.4$
\\
\hline
\end{tabular}
\end{table}

\subsection{Structure inversion}
\label{subsec:numerical_tests_for_sturc_inv}

\begin{figure*}
	\includegraphics[width=\columnwidth]{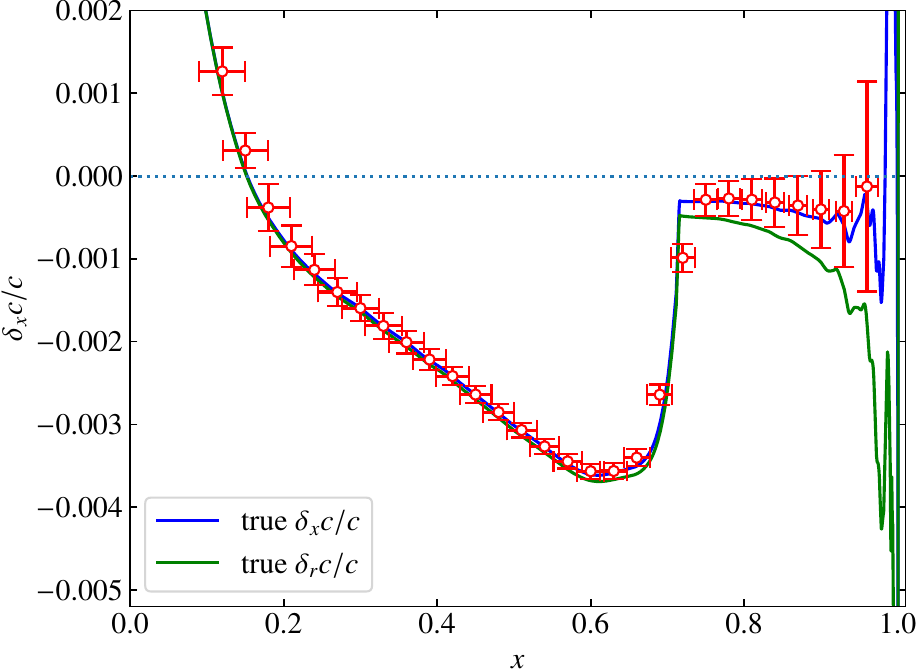}
 	\includegraphics[width=\columnwidth]{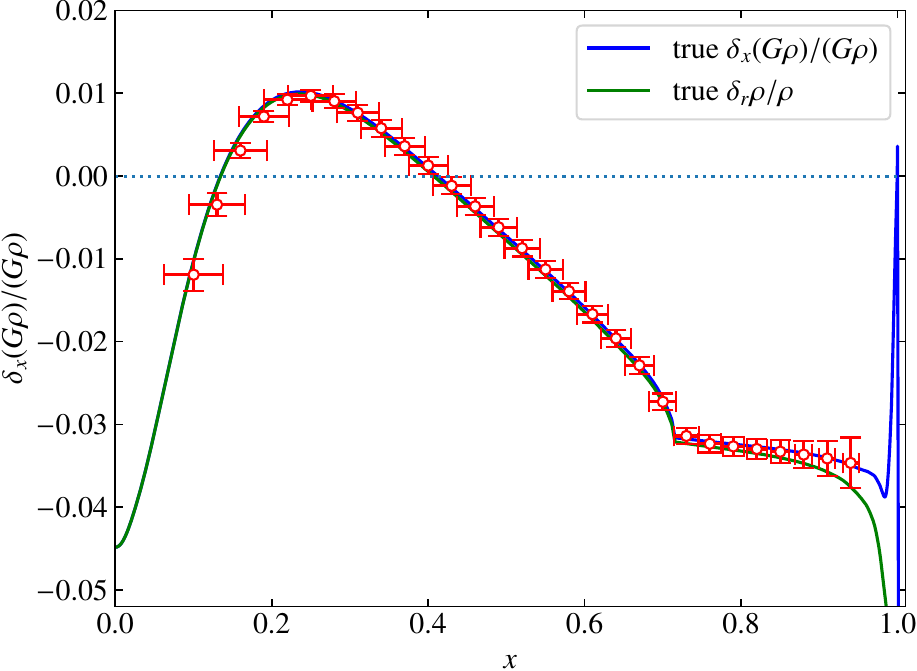}
    \caption{%
    {%
    Test} structure inversions based on two theoretical models
    (cf.~{Section} \ref{sec:revision}).
    One is
    {model} S of \citet{Christensen-Dalsgaard:1996aa},
    which is used as
    the reference model,
    while
    the other is
    {model} 1 of \citet{HGseconddiff2007MNRAS},
    which is homologously shrunk by
    $0.01$~{per cent}.
    The inversions for $\delta_x c/c$
    and $\delta_x (G\rho)/(G\rho)$
    are shown 
    in the left and right panels,
    respectively,
    {
    as functions of the fractional radius $x$
    of the reference model,
    which is normalized by
    the photospheric radius.}
    In each panel,
    the red open circles
    with errorbars
    indicate the inversion results.
    The horizontal bars stand for
    the width of the averaging kernels,
    while the vertical errorbars stand for
    the statistical errors that originate from
    the uncertainties in frequencies,
    which are assumed to be the same
    as those of MDI 360-day data set.
    The blue curves 
    {denote} the corresponding true differences;
    the green curves, which
    {represent}
    $\delta_r c/c$ and $\delta_r\rho/\rho$
    in the left and right panels, respectively,
    are {included} for comparison.
    }
    \label{fig:struc_inv_test}
\end{figure*}

\begin{figure*}
	\includegraphics[width=\columnwidth]{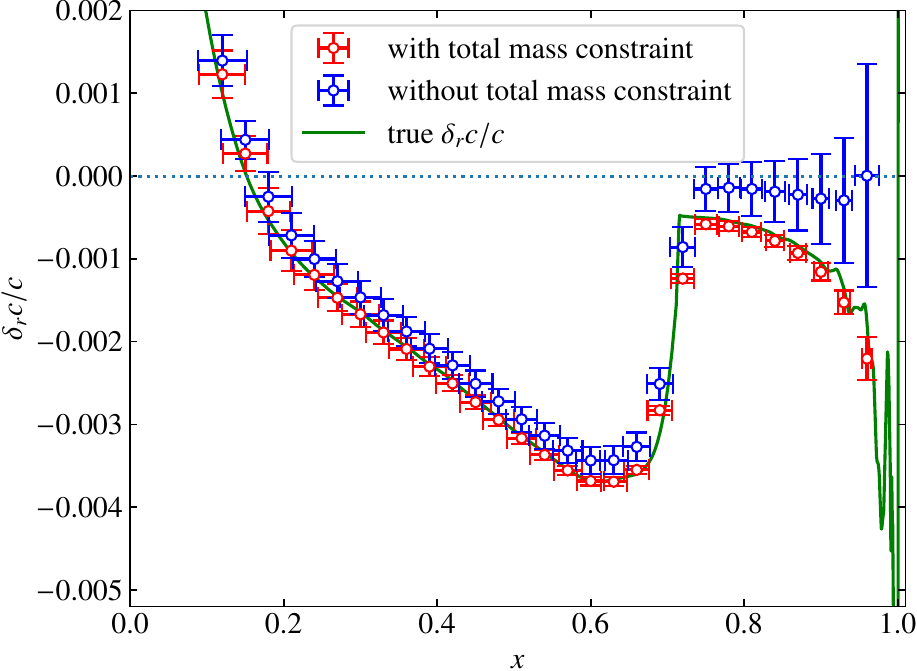}
 	\includegraphics[width=\columnwidth]{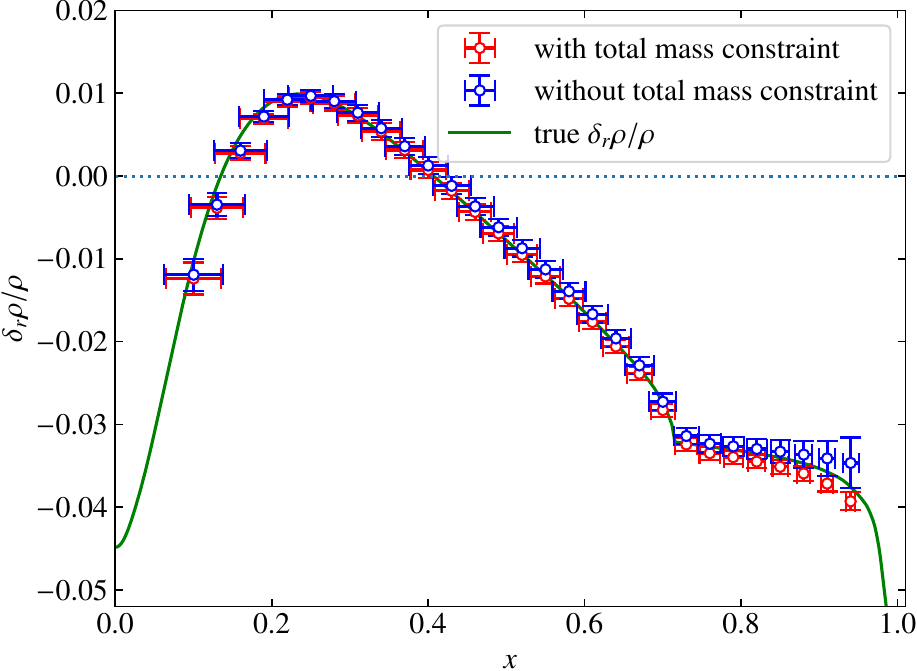}
    \caption{%
    Test structure inversions by the conventional method,
    which does not consider
    the difference between 
    the radii of
    target structure and the reference model.
    The target and reference models are the same as those in Fig.~\ref{fig:struc_inv_test}.
    The left and right panels show
    the sound speed and density differences,
    respectively, at fixed radius.
    The inversion results 
    with (without)
    the total mass constraint
    are plotted by
    red (blue) open circles with  errorbars,
    and the corresponding true differences are
    drawn by green curves.
    {
    The meanings of the horizontal axes
    and the errorbars follow Fig.~\ref{fig:struc_inv_test}.
    }
    }
    \label{fig:conv_struc_inv_test}
\end{figure*}

{
We validate
the method of structure inversion,
which is described in {Sections}
\ref{subsec:mod_c_inv}
and
\ref{subsec:mod_rho_inv},
based on the same theoretical models
as those in {Section} \ref{subsec:numerical_tests_for_R_inv}.
We use all the p modes in the dataset,
which is common with the MDI 360-day data,
but exclude f modes.
}
{
Fig.~\ref{fig:struc_inv_test}
shows the results of the 
inversions
for $\delta_x c/c$ (left panel)
and $\delta_x (G\rho)/(G\rho)$ (right panel).
{They} are fully consistent with the true differences (drawn by blue curves) in the both cases.
}
{
In the left panel, 
we observe that
the values of $\delta_x c/c$
are nearly constant
in the convective envelope
($x \gtrsim 0.7$),
whereas the true difference
of $\delta_r c/c$
(green curve)
increases in magnitude towards the surface.
This is the direct effect of scaling
introduced in the inversion process.
In fact,
we understand from equation
(\ref{eq:drdx_dif2})
}
\begin{equation}
\frac{\delta_x c}{c}
=
\frac{\delta_r c}{c}
+
\frac{\delta R}{R}
\frac{\rd\ln c}{\rd\ln r}
\;.
\label{eq:dxlnc_drlnc}
\end{equation}
The almost constant {small} 
values of $\delta_x c/c$ 
{in the convection zone beneath the 
superadiabatic boundary layer} 
implies
that the second term
on the right-hand side of equation (\ref{eq:dxlnc_drlnc}) 
{nearly cancels 
the first term (the green curve
in Fig.~\ref{fig:struc_inv_test})}.
The corresponding difference
in the density inversions
can be understood 
{similarly.}

{
We then check any potential troubles in
the conventional inversion method.
In Fig.~\ref{fig:conv_struc_inv_test},
we present the inversion results
for $\delta_r c/c$ (left panel)
and $\delta_r \rho/\rho$ (right panel)
based on the conventional method.
We assume that
the target structure 
(shrunk {model} 1)
has the same radius as the reference model
({model} S).
Two cases with and without
the total mass constraint
are shown in each panel.
The results for $\delta_r c/c$
with the total mass constraint
(red points in the left panel)
are consistent with the true difference (green curve).
On the other hand,
those without the total mass constraint (blue points) are systematically larger by
$\sim 10^{-4}$ than
the true curve (green)
for $x \lesssim {0.7}$,
whereas 
the difference becomes larger
as $x$ increases
for $x \gtrsim {0.7}$,
and reaches $\sim 2\times 10^{-3}$
for $x = 0.96$.
These {values} are formally
obtained by shifting upward
the red points in the left panel
of Fig.~\ref{fig:struc_inv_test}
by $10^{-4}$.
}
{This is because
the conventional inversion for $\delta_r c/c$
without the total mass constraint
should be reinterpreted as
$\delta_x \covr/\covr
= \delta_x c/c - \delta R/R$}
(cf.~{Section} \ref{subsec:annihilator_relation}).

{
The  conventional inversions 
for $\delta_r\rho/\rho$ 
with the total mass constraint
(red points in the right panel)
are systematically smaller 
{than the true values (green curve)}
by
$\sim 10^{-3}$ for $x \gtrsim 0.7$.
These differences could be understood  
{as} 
a nonlinear effect, which we may estimate to be of the order of $(\delta_r\rho/\rho)^2 \sim (3\times 10^{-2})^2 \sim 10^{-3}$ 
in the relevant range of $x$.
On the contrary,
the results without the total mass constraint
(blue points) are larger than
the true curve (green)
by $\sim 10^{-3}$ for $x \gtrsim 0.8$.
These results are numerically
the same as those
shown by red open circles
in the right panel of
Fig.~\ref{fig:struc_inv_test}
{since
$\delta_r \rho/\rho$
without the total mass constraint
should be reinterpreted as
$\delta_x(G\rho)/(G\rho)$
(cf.~{Section} \ref{subsec:annihilator_relation})
}.
}

\section{Inversion of observational data}
\label{sec:real_R_inv}

{In this section,
we apply the inversion methods
developed in Sections
\ref{subsec:OLA_inv_dR},
\ref{subsec:mod_c_inv}
and 
\ref{subsec:mod_rho_inv},
for the radius, sound speed and density,
respectively,
to observational data.
The frequency data set used in this paper was obtained from
the SOI/MDI instrument on the SOHO spacecraft
\citep{Schou:1999aa}.
We {use} 
all
the 
{p}
modes included in the 
available
360-day data.
{Again} we adopt model S by 
\citet{Christensen-Dalsgaard:1996aa} as the reference 
model.}

{
We {do} not {include} f modes in the analysis. 
This is chiefly because
p modes and f modes
are sensitive to
different radii
as we discuss in {Section} \ref{sec:seismic_radii}.
{In addition, a}
complication arises from the 
fact that
the surface terms
for 
the two kinds of mode differ,
because the characteristics of 
the  modes are rather different, 
partly because 
f modes are uncompressed
{(cf.~{Section} \ref{subsec:OLA_inv_dR})}.
Indeed,
the functions $F_0$ and $F_2$ for p modes
are 
inapplicable 
to f modes, 
because f-mode frequencies are intimately related to degree.
}

\subsection{Structure inversion}
\label{subsec:struc_inv_obs}

\begin{figure*}
\centering
\includegraphics[width=\columnwidth]{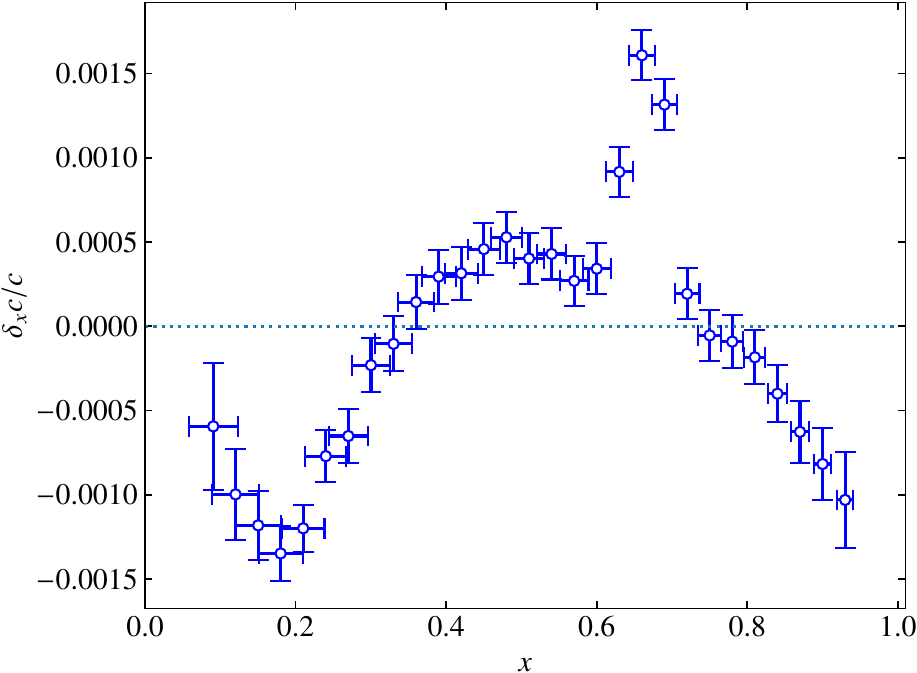}
\includegraphics[width=\columnwidth]{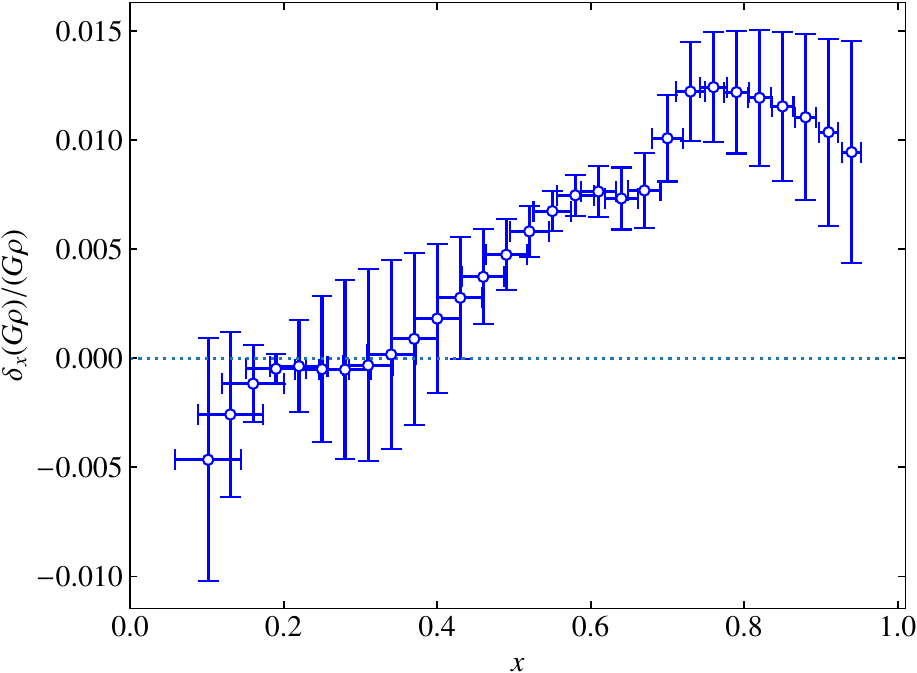}
\caption{%
Inversions for the structure of the Sun
based on the MDI 360-day data.
The method is described
in detail in {Section}~\ref{sec:revision}.
The results for the sound speed and the density multiplied by the gravitational constant are given
in the left and right panels, respectively.
The ordinate of each panel means
the relative difference 
of the solar values
from those of the reference model
at the fixed fractional radius, $x$.
{
    The meanings of the horizontal  {bars} 
    and the {vertical} errorbars follow Fig.~\ref{fig:struc_inv_test}.
    }
}
\label{fig:conventional_vs_new}
\end{figure*}

%
%
%
In the left panel of Fig.~\ref{fig:conventional_vs_new},  
sound-speed inversions obtained by 
the method described in {Section}~\ref{subsec:mod_c_inv}
are presented.
Since the claimed
relative difference between the radii of the Sun and the
standard solar model is of the order of $10^{-4}$
\citep{Schou:1997aa,Antia:1998aa},
its second-order effect on the new structure inversions is
expected to be of the order of $10^{-8}$,
which is small enough to be safely ignored at the current level
of the accuracy of
the eigenfrequency measurements.
{
Unlike in the case of the test inversions
in {Section} \ref{subsec:numerical_tests_for_sturc_inv},
$\delta_x c/c$ in the convective envelope
increases in magnitude towards the surface.
This implies
that this region
can be described by different scaling
from what is adopted in these inversions,
as we discuss in {Section}
\ref{subsec:c_in_CE}.
}

{The corresponding
density inversions are
depicted
in the right panel of Fig.~\ref{fig:conventional_vs_new}.
We observe that
the values of $\delta_x (G\rho)/(G\rho)$ are
mostly positive.
This is understandable because
we do not use the total mass constraint.
As in the case of the sound-speed inversions,
the values of $\delta_x(G\rho)/(G\rho)$ in the convective
envelope tend to decrease towards the surface.
}




\begin{table}
\caption{
Radius inversions using only the p modes.  $\delta R$ is the estimated 
radius scale of the Sun minus the radius scale of the reference, {model} S. 
Parameters $n_0$ and $n_2$ denote the numbers of terms in the expansions
of the surface-term functions $F_0$ and $F_2$
in equation (\protect\ref{eq:surface_term}), respectively;
{$\alpha_R$, $\beta_R$ and $\gamma_R$ are parameters in the formula (\ref{eq:chi2_R}) 
for $\chi^2_R$ whose minimization determines the coefficients $c_{n,l}$;
{in all the cases they are fixed 
at $\alpha_R = 1$, $\beta_R = 10^2$ and $\gamma_R = 5\times 10^7$.}}
In the inversions, it was assumed that $\delta (GM)/(GM) = 0$ 
 as described after equation (\ref{eq:dR_estimate2}),
 because the relative error in the measurement of $\GMsun$ is
 on the order of $10^{-11}$ (see Table \ref{tab:global_quantities}).
 An estimate of the error in 
${(}\delta R/R{)_{\text{ac}}}$ is provided by $\sigma$; its value is rather `large' as a result of a trade-off with the suppression 
of the contaminating integrals {of equation (\ref{eq:C_def})} listed in the last two columns. 
{
Here, we define the cross-talk integrals
$C_{\covr}:=
\int \mathcal{K}^{{(R)}}_{\covr,\rho}\; \delta_x \covr/\covr \dx$
and
$C_{\rho}:=
\int \mathcal{K}^{{(R)}}_{\rho,\covr}\; \delta_x (G\rho)/(G\rho) \dx$.
}
}
\label{tab:R_surface_term_effect}
\centering
{{$\begin{array}{cccccc}
\hline
n_0 & n_2 & (\delta R/R)_{\text{ac}} & \sigma
  & C_{\covr} & C_{\rho}
\\
\hline
%
12 &  0 & -4.3\times 10^{-4}  & 6\times 10^{-5}
  & -6\times 10^{-5} & -7\times 10^{-5}\\
12 & 12 & -3.1\times 10^{-4}  & 7\times 10^{-5}
  & -6\times 10^{-5} & -6\times 10^{-5}\\
20 & 20 & -3.5\times 10^{-4}  & 7\times 10^{-5}
  & -6\times 10^{-5} & -5\times 10^{-5}\\
40 & 40 & -3.1\times 10^{-4}  & 8\times 10^{-5}
  & -7\times 10^{-5} & -5\times 10^{-6}\\
\hline
 \end{array}
$}}
\end{table}

\subsection{Radius inversion}

{We perform
the radius inversion
following the method in
Section \ref{subsec:OLA_inv_dR}.}
Although the absolute values of 
the two integrals
in equation (\ref{eq:C_def})
should be small enough not to affect the inversions,
these terms could be sources of systematic error
in the final answers given by equation (\ref{eq:dR_estimate2}).
We estimate 
these integrals
using the profiles of 
{$\delta_x\covr/\covr$
and $\delta_x(G\rho)/(G\rho)$}
obtained in
the inversion
without the total mass constraint
{(cf. }{Section} 
\ref{sec:revision}{)}.
{The results are given in Table
\ref{tab:R_surface_term_effect}.}

In {Table} \ref{tab:R_surface_term_effect},
we {also} check the sensitivity of the inversions to
the numbers of terms included in the expansion of the surface-term
functions $F_0$ and $F_2$, which we denote by $n_0$ and $n_2$, respectively.
We see that the results are insensitive to both $n_0$ and $n_2$,
provided they both exceed 12.
%
We adopt the last entry in {Table} \ref{tab:R_surface_term_effect} as the
final answer of the present study.
To be conservative, we regard the 3-$\sigma$ level of
the formal error as the uncertainty in the current estimate of the radius
difference between the Sun and the reference model.
Hence we estimate that 
{ $(\delta R/R)_{\text{ac}} = (-3.1 \pm 2.4) \times 10^{-4}${.}}
{
Combining this with
equations
(\ref{eq:scaling_assumption})
and
(\ref{eq:dRac_dRc}),
we obtain 
the p-scaled (photospheric) radius as
$\Rp=\Rpval$.
}
{Recall that, strictly speaking, this result depends  on assuming a}
homologous difference
in the structure of
the outer layers 
of the Sun
beyond the upper turning point
of the p modes included in the present study.
{Here, we
have used 
the photospheric radius of
the reference model,
${\Rphr=} 695.99~\text{Mm}$.}


\section{Discussion}\label{sec:discussion}



{Although}
\citet{Schou:1997aa} 
{analysed f-mode frequencies to}
give
${\Rf}=695.68\pm 0.03$~Mm,
{the errors are dominated by
the systematic errors.
The corresponding
3-$\sigma$ statistical errors are
estimated to be
$0.02$~Mm.
}
Although their estimate has a smaller formal error than ours,
their result may be sensitive to the description of the subsurface layer of
superadiabatic convection, as they point out themselves.
On the other hand,
our analysis is expected to be almost free from that ambiguity,
partly because we have used p modes which are evanescent in the most
turbulent region of the convection zone and partly
because we have taken account of as many as 40 terms in the expansion of the
surface term, which
is expected to remove
the uncertainty concerning the subsurface structure
to a considerable extent.
\citet{Dziembowski:2000aa} also analysed f-mode data,
implying
{the relative difference between
the f-mode radii of
the Sun and {model} S \citep{Christensen-Dalsgaard:1996aa} of}
{${\delta\Rf/\Rf}
=(-4.52\pm 0.03)\times 10^{-4}$}
($\pm$ one standard deviation),
averaged over nearly 3 years;
the major variation in {$\Rf$} appears to be an oscillation with a 1-year
period,
which may suggest a susceptibility of the analysis
at the $10^{-5}$ level
to an annual variation in the SOHO--Sun distance
resulting from, for example,
pixel quantization
or
instrumental temperature variations,
{
not to mention genuine
solar-cycle radius variation of the Sun.
}

Alternatively,
\citet{Antia:2003aa} claims that
the apparent time-variation of the 
{f-scaled} radius
could 
{have been caused}
by a change in the MDI/SOI instrument
during the few-months `vacation' of the SOHO spacecraft
in 1998 and 1999.
{Subsequently,
\citet{Lefebvre:2005aa} and
\citet{Lefebvre:2007aa}
studied the nonhomologous solar-cycle variation
of the subsurface layers
based on the f-mode frequencies.
\citet{Rozelot:2018aa}
have reported another interesting
result:
the f-mode frequencies 
are correlated with
sunspot numbers
over nearly two cycles ($\sim 22$ years).
}


We find that
the central value of our radius inversion lies between
the two
direct observations quoted by
\citet{Allen:1973aa} and \citet{Brown:1998aa}, 
though 
{the former of them is}
consistent with our result
at the 3-$\sigma$ level of the error.

\subsection{Sound-speed inversions in the convective envelope}
\label{subsec:c_in_CE}

We observe
in the left panel of Fig.~\ref{fig:conventional_vs_new}
that
the sound-speed inversions for $x\gtrsim 0.7$
decrease monotonically towards the surface.
It is well established that this region is composed of
essentially adiabatically stratified layers,
in which $p = K \rho^{\gamma_1}$ 
with constant $K$.
The sound speed is then approximately described by
{%
\begin{equation}
c^2 \approx \left(\gamma_1 - 1\right) \GMsun
\left( \frac{1}{r} - \frac{1}{\Rs} \right)
\;,
\label{eq:csq_conv}
\end{equation}
}%
which can be derived from the equations of
hydrostatic equilibrium
{under the assumptions of
constant $\gamma_1$ and {constant} mass 
{enclosed within radius $r$.}}
Here $\Rs$ is the location of the phantom singularity 
that was introduced in Section \ref{sec:seismic_radii}.
The expression for $\delta_x c/{c}$ can be derived from 
equation (\ref{eq:csq_conv}) as 
\begin{equation}
\frac{\delta_x c}{c}
\approx
\frac{1}{2}
\left(
{
\frac{\delta\gamma_1}{\gamma_1 - 1}
}
-
\frac{\delta R}{R}
\right)
-
\frac{1}{2\Rs}
\left(
\frac{\delta R}{R} - \frac{\delta \Rs}{\Rs}
\right)
\left(
\frac{1}{r}-\frac{1}{\Rs}
\right)^{-1}
\;.
\label{eq:dxlnc_CE}
\end{equation}
The growing trend in the sound-speed inversions
can be interpreted as
a contribution from the second term
on the right-hand side of equation
(\ref{eq:dxlnc_CE}).
{
Comparing
the approximate relation (\ref{eq:dxlnc_CE})
with
a more accurate expression,
\begin{equation}
\frac{\delta_x c}{c}
=
\frac{\delta_{\xs} c}{c}
+
\left(
\frac{\delta R}{R}
-
\frac{\delta\Rs}{\Rs}
\right)
\frac{\rd\ln c}{\rd\ln r}
\;,
\label{eq:dlnc_acc}
\end{equation}
in which $\xs = r/\Rs$,
we may regard 
the first term on the right-hand side of
equation (\ref{eq:dlnc_acc}) {as being}  
almost constant{, as one might expect}.
Fitting equation (\ref{eq:dlnc_acc})
to the sound-speed inversion
performed in {Section} \ref{subsec:struc_inv_obs},
}
{we} estimate{, in particular,} 
\begin{equation}
\frac{\delta \Rs}{\Rs} - 
\left(\frac{\delta R}{R}\right)_{\text{ac}}
\approx {-0.0003}
\;.
\end{equation}
This means that
the position of the adiabatically stratified layers,
which is characterized by $\Rs$,
is located about {$0.06$~{per cent}} deeper in the Sun
than in {model} S,
whereas $\Rc$,
which is almost equal to the photospheric radius,
is smaller
by only $0.03$~{per cent}.
The result of $\delta \Rs/\Rs \ne 
{(}
\delta R/R
{)_{\text{ac}}}$
means that
the difference between the Sun and
{model} S is not homologous
for $x \gtrsim 0.7$.
This information would be useful
for our better understanding
of the structure of the upper convective layers
in the Sun.

\subsection{Remark on \texorpdfstring{\citeauthor{Basu:1998aa}}{Basu}}

\citet{Basu:1998aa} performed {conventional} structure inversions
using two different reference models, one with the standard value of the 
photospheric radius,
$695.99$~Mm,
the other with the smaller radius 
${\Rph}=695.78$~Mm.
She found a significant difference between the results,
both in the sound-speed and in the density inversions.
Her demonstrations emphasize that 
so long as we adhere to conventional 
{inversions}
we must 
{take care in
interpreting the results.}
On the other hand,
by extending the inversion formulae so as to
take account of the uncertainties in the solar radius $\Rsun$
{and}
the product $\GMsun$ of the gravitational constant and the solar mass,
we have succeeded in 
carrying out 
inversions that are independent of the
radius differences (at least in the leading order).
{%
Note that the uncertainty in $\GMsun$
can be neglected in practice because
it is much smaller than
the errors in frequencies
provided by current observations.}
We can say that our method gives conservative answers,
in the sense that
it makes the results independent of the uncertainties in the radius $R$
at the expense of greater formal errors.


\subsection{{Uncertainty in the gravitational constant}}

In principle, the  density inversions are affected by
any error $\sigma_G$ 
in the gravitational constant, which is of the order of
$10^{-5}$
(cf.~{Table}~\ref{tab:global_quantities}), although the formal errors
in the right panel of Fig. \ref{fig:conventional_vs_new}
are larger by about
{two orders}
of magnitude.
Similarly, if there were a difference in the gravitational constant $G$
between the Sun and the reference model, then the density inversions
should be shifted by a constant $\delta G/G$ since
\begin{equation}
 \frac{\delta_x \rho}{\rho}
=
 \frac{\delta_x (G\rho)}{G\rho}
-
 \frac{\delta G}{G}
\;.
\end{equation}
%
%

\subsection{Relation to asteroseismology}

It is worth thinking about the application of the present technique
to the field of asteroseismology
\citep[cf.][]{Gough:1993ab,Reese:2012aa,Buldgenetal2019MNRAS.482.2305B}.
The present method 
enables us to perform the structure inversions that are 
independent of the uncertainties in the total radius $R$
if we know the product $GM$ of the gravitational constant and the total mass 
of the target star accurately.
If $GM$ were not accurately known,
but the radius $R$ of the star were measured,
for example,
by the interferometric observations
\citep[e.g.][]{Kervella:2004aa},
it would   
still be possible to perform the inversions
for the structure difference
and the total mass difference
(if frequencies were available for a large variety of oscillation modes).
In fact,
the inversions 
for $\delta_x \covr/\covr$ and $\delta_x (G\rho)/(G\rho)$
can be accomplished using only equation
(\ref{eq:perturbed_VP}), 
as is explained in
{Section} \ref{subsec:radius_determination}.
In this case,
we do not have to worry about
any ambiguity in
the definition of the operator $\delta_x$
since the radius difference is known.
Then
the mass difference $\delta (GM)/(GM)$
can be inferred by
another OLA-type inversion
that  utilizes
the total mass constraint
given by equation
(\ref{eq:mass_constraint}).
If, in the worst case, we know neither 
$R$ nor the product
$GM$ of the target star,
we 
can 
estimate only the difference
in $GM/R^3$
by a procedure similar to that described above.
A reliable estimate of the 
surface gravity $GM/R^2$ by spectroscopy
(or the theoretical mass-radius relation) then 
constrains both of the radius and the mass of the target star.
{
These days,
the mass and radius of solar-like oscillators
are often estimated
from the large frequency separation, $\Delta\nu$, and 
the frequency of maximum power,
$\nu_{\max}$,
based on
the scaling relations.}%

Once the radius difference is known,
the structure differences
$\delta_x \covr/\covr$ and $\delta_x (G\rho)/(G\rho)$,
which are inferred by equation (\ref{eq:perturbed_VP}),
can be defined without ambiguity.




 
\subsection{The case of the large radius difference}

We note in passing
that
the assumption that
$\delta R/R$ is small
is not actually essential,
though it is
implicitly made
when we write down
equation
(\ref{eq:drdx_dif2}).
In fact,
we can {carry out a} similar  {analysis} 
even {when} $\delta R/R$ is not small,
{%
provided that
the structure difference is nearly homologous.
Here the nearly homologous difference means
that 
the differences at fixed fractional radius in
the dimensionless variables
\begin{equation}
\tilde{\covr} := \sqrt{\frac{R^3}{GM}}\, \covr
\end{equation}
and
\begin{equation}
\tilde{\rho} := \frac{R^3}{M}\, \rho
\end{equation}
can be made small
by adjusting the radius (scale factor)
of the target structure.
Let us denote the radii $R$
of the reference model and the target structure
by $\Rr$ and $\Rt$, respectively.
In this section,
subscripts $\text{r}$ and $\text{t}$ generally mean
the quantities
of the reference model and the target structure,
respectively.
We do not assume
$\left|\Rt - \Rr\right| \ll \Rr$.
We first set $\Rt = \Rtz$, for which
the corresponding difference operator
at fixed radius is denoted by $\delta_{x,0}$.  
The value of $\Rtz$ can be chosen arbitrary
so long as $\delta_{x,0}\tilde{\covr}$ and $\delta_{x,0}\tilde{\rho}$ are small.
We then consider variation in $\Rt$ as
$\Rt = \Rtz + \delta \Rt$,
in which we assume
$\left|\delta \Rt\right| \ll \Rtz$.
The corresponding difference operator at fixed fractional radius
is denoted by $\delta_{x}$.
{After} these preparations,
we may repeat
{the analysis} in 
{Section} \ref{015032_29Oct18}, {but} 
with 
$\covr$, $G\rho$, $\delta R/R$ and
$\delta_r$
{replaced}
by
$\tilde{\covr}$, $\tilde{\rho}$,
$\delta \Rt/\Rtz$ and
$\delta_{x,0}$.
Equation (\ref{eq:dRR_rho_inf}){, in particular, is}
reduced to
\begin{equation}
\frac{\delta \Rt}{\Rtz}
=
\lim_{x_0\rightarrow\xsurf}  
\frac{\displaystyle
\frac{\delta_{x,0} \tilde{\rho}}{\tilde{\rho}_{\text{r}}}(x_0)}
{\displaystyle
 -\frac{{\rd}\ln\rho_{\text{r}}}{{\rd}\ln r}(x_0)}
 \;,
\end{equation}
which is equivalent to
\begin{equation}
\lim_{x_0\rightarrow\xsurf}  
\frac{\displaystyle
\frac{\delta_{x} \tilde{\rho}}{\tilde{\rho}_{\text{r}}}(x_0)}
{\displaystyle
 -\frac{{\rd}\ln\rho_{\text{r}}}{{\rd}\ln r}(x_0)}
 = 0
 \;.
 \label{eq:dRr_general}
\end{equation}
}%
We can easily understand
how 
the scale factor {$\Rt$}
is included in
{equation (\ref{eq:dRr_general})}
by remembering that
{
\begin{equation}
\delta_x \tilde{\rho} (x)
=
\frac{\Rt^3}{\Mt}\,
\rho_{\text{t}}\left(x \Rt\right)
-
\frac{\Rr^3}{\Mr}\,
\rho_{\text{r}}\left(x \Rr\right)
\;.
\end{equation}
The scale factor $\Rt$,
which can be very different from $\Rr$,}
is identified
as the zero point of
 the left-hand side of equation
(\ref{eq:dRr_general}){, regarded as a function of $\Rt$.}
{That, of course, is provided that the target 
and the reference have essentially the same outer 
atmospheres, as is
the case for the models studied in 
{Section} \ref{subsec:numerical_tests_for_R_form}.}  
{This} analysis{, which} 
allows {for a} 
large radius difference{,} 
must
be of use
when we think about
inversion{s} of stars other than the Sun,
whose radii we often do not know well.


\section{Conclusion}\label{sec:conclusion}

In this paper, 
we have extended the inversion formulae to
consider
the
difference in the radii of the reference model and the Sun{.}
{We} have performed inversions for the radius of the Sun,
and estimate the solar photospheric radius (which we quote with 3-$\sigma$ statistical errors) to be
$\Rpval$ from only p-mode frequencies.
{%
We have also performed structure inversions
for the sound-speed and density profiles of the Sun
independently of the uncertainties in the solar radius.
The {sound-speed} inversion suggests
that the positions of the photosphere and the adiabatically stratified layers in the convective envelope
differ nonhomologously from those of the standard solar model.
}

\section*{Acknowledgements}
We thank J.~Schou for providing us with the SOHO/MDI frequency data set
used for the inversions.
T.~Sekii 
is thanked for
insightful discussions
and
providing inversion programmes,
which have been extended for the present study.
MT
gratefully
acknowledges
the helpful comments of
H.~Shibahashi;
{he} is {also} grateful to PPARC (UK) for financial support.
This work
was 
also
supported by 
JSPS KAKENHI Grant Numbers 
JP12047208,
JP26400219
and 
JP18K03695
and
a Japan-UK Joint Research project
of
the Japan Society for the Promotion of Science (JSPS).
DOG is grateful to JSPS
for 
an Invitation Fellowship, and to the Leverhulme Trust for an Emeritus 
Fellowship.  {We thank J. Christensen-Dalsgaard for useful
conversations, and}
{our referee for meticulously commenting 
on the manuscript.}

\section*{Data availability}
The data underlying this article will be shared on reasonable request to the corresponding author. 



\bibliographystyle{mnras}
\bibliography{references}


\appendix


\section[]
{Derivation of the inversion formulae}\label{sec:derive_pt_VP}

Here we derive equation (\ref{eq:perturbed_VP}) 
for the 
fractional difference $\delta \nu_{n,l}/\nu_{n,l}$ between the solar 
frequencies and the eigenfrequencies of a reference model in terms of the 
corresponding structural differences in sound speed and density.
First,
we introduce the dimensionless variables,
\begin{align}
 \tilde{\nu}_{n,l} & := \sqrt{\frac{R^3}{GM}}\; \nu_{n,l} \;, 
\label{eq:nondim_nu}\\
 \tilde{c}     & := \sqrt{\frac{R}{GM}}\; c
\label{eq:nondim_c}
\\\noalign{\noindent and}
 \tilde{\rho}  & := \frac{R^3}{M}\; \rho \; .
\label{eq:nondim_rho}
\end{align}
The equations of hydrostatic equilibrium and mass conservation
can be rewritten in terms of these dimensionless variables, 
and are 
quite similar to the original forms. 
With these equations,
the derivation of the kernels for
dimensionless 
{sound-speed} $\tilde{c}$
and dimensionless density $\tilde{\rho}$
is parallel to that for the dimensional variables,
for which
readers can refer, for example, to the article by \citet{Gough:1991aa}.
The difference is 
that we should compare the variables
at the fixed fractional radius $x=r/R$ instead of the fixed absolute radius $r$.
The resulting integral expression is  
\begin{equation}
\frac{\delta\tilde{\nu}_{n,l}} {\tilde{\nu}_{n,l}}
=
\int K_{\tilde{c},\tilde{\rho}}^{(n,l)} \,
 \frac{\delta_x \tilde{c}}{\tilde{c}} \dx
+
\int K_{\tilde{\rho},\tilde{c}}^{(n,l)} \,
 \frac{\delta_x \tilde{\rho}}{\tilde{\rho}} \dx
+
S_{n,l}
\;,
\label{eq:nondim_perturbed_VP}
\end{equation}
where $S_{n,l}$ is 
introduced
to accommodate the uncertain 
physics in the near-surface layers of the star. 
{It is therefore only very weakly dependent on the degree 
$l$ of the mode.}
{The reason for this is because
the ray paths of the high-order p modes are nearly vertical
in the near-surface layers, irrespective of degree.}
Noting 
that the kernels for the dimensional variables are dimensionless,
we can easily appreciate that
\begin{equation}
K_{\tilde{c},\tilde{\rho}}^{(n,l)} = K_{c,\rho}^{(n,l)}\;\;\;\;\;{\text{and}}\;\;\;\;\;
K_{\tilde{\rho},\tilde{c}}^{(n,l)} = K_{\rho,c}^{(n,l)}:
\label{eq:Kcandrho_nondim=dim}
\end{equation}
the kernels for the dimensionless variables are
identical to those for the dimensional variables.
Note that equation (\ref{eq:nondim_perturbed_VP}) holds
even for the case in which
the real Sun and the reference model have different radii.


We next rewrite equation (\ref{eq:nondim_perturbed_VP})
in terms of the dimensional variables
by using the definitions
(\ref{eq:nondim_nu}), (\ref{eq:nondim_c}) and (\ref{eq:nondim_rho}):
\begin{align}
\displaystyle
 \frac{\delta\tilde{\nu}_{n,l}}{\tilde{\nu}_{n,l}} & =
 \frac{\delta\nu_{n,l}}{\nu_{n,l}} 
 + \frac{3}{2} \, \frac{\delta R}{R}
 - \frac{1}{2} \, \frac{\delta (GM)}{GM} \;,
\label{eq:nondim_dnu}
\\
\displaystyle
 \frac{\delta_x \tilde{c}}{\tilde{c}} & =
 \frac{\delta_x c}{c}
 + \frac{1}{2} \, \frac{\delta R}{R}
 - \frac{1}{2} \, \frac{\delta (GM)}{GM}
\label{eq:nondim_dc}
\\\noalign{\noindent and}
\displaystyle
 \frac{\delta_x \tilde{\rho}}{\tilde{\rho}} & =
 \frac{\delta_x (G \rho)}{G \rho}
 + 3 \frac{\delta R}{R}
 - \frac{\delta (GM)}{GM} \;.
\label{eq:nondim_drho}
\end{align}
Introducing
these relations
together with 
equation
(\ref{eq:Kcandrho_nondim=dim})
into equation
(\ref{eq:nondim_perturbed_VP}) 
yields 
\begin{align}
\displaystyle
 \frac{\delta\nu_{n,l}}{\nu_{n,l}} = &
 \int K_{c,\rho}^{(n,l)} \, \frac{\delta_x c}{c} \dx
+
 \int K_{\rho,c}^{(n,l)} \, \frac{\delta_x (G \rho)}{G \rho} \dx
\nonumber \\
& \displaystyle \mbox{}
+ \frac{\delta R}{R}
  \left(
  \frac{1}{2} \int K_{c,\rho}^{(n,l)} \dx + 3 \int K_{\rho,c}^{(n,l)} \dx
 - \frac{3}{2}
  \right)
\nonumber \\
& \displaystyle \mbox{}
- \frac{\delta (GM)}{GM}
  \left(
  \frac{1}{2} \int K_{c,\rho}^{(n,l)} \dx + \int K_{\rho,c}^{(n,l)} \dx - \frac{1}{2}
  \right) 
\nonumber \\
& \displaystyle \mbox{}
+
S_{n,l}
 \;.
\label{eq:dim_perturbed_VP_var}
\end{align}
We can go further by comparing
equation (\ref{eq:dim_perturbed_VP_var})
with the corresponding formula used for the conventional inversion,
which is a particular case of equation (\ref{eq:nondim_perturbed_VP}):
\begin{equation}
 \frac{\delta\nu_{n,l}}{\nu_{n,l}} =
 \int K_{c,\rho}^{(n,l)} \, \frac{\delta_r c}{c} \dx
 +
 \int K_{\rho,c}^{(n,l)} \, \frac{\delta_r \rho}{\rho} \dx
 +
 S_{n,l}
 \;.
\label{eq:conv_perturbed_VP}
\end{equation}
We 
now emphasize 
that
we must be able to recover equation (\ref{eq:conv_perturbed_VP})
from
equation (\ref{eq:dim_perturbed_VP_var}),
which is more general.
To do so,
all we need should be only to set
$\delta G = 0$ and $\delta R = 0$ (hence $\delta_x = \delta_r$), 
because
equation (\ref{eq:conv_perturbed_VP}) has been derived
without assuming $\delta M = 0$.
%
However, it is found that
the substitution of those
two conditions
into
equation (\ref{eq:dim_perturbed_VP_var}),
and
the replacement of $\delta_x$ by $\delta_r$, 
are not sufficient to recover
equation (\ref{eq:conv_perturbed_VP})
unless
the terms multiplying ${\delta(GM)}/{GM}$ cancel.
%
In other words,  the identity 
\begin{equation}
\frac{1}{2} \int K_{c,\rho}^{(n,l)} \dx + \int K_{\rho,c}^{(n,l)} \dx = \frac{1}{2}
\;,
\label{eq:it_identity}
\end{equation}
must be satisfied for all $(n,l)$.
From a physical point of view,
equation (\ref{eq:it_identity}) reflects
the following homologous relation of the adiabatic oscillations of 
stars:
if we multiply both 
the squared sound-speed profile and the density profile of a {stellar model} 
by the same constant factor, 
keeping their shapes fixed as functions of the fractional radius $x$,
all of the squared eigenfrequencies change by the same factor.
Using identity (\ref{eq:it_identity}),
%
equation (\ref{eq:dim_perturbed_VP_var}) 
can be written 
\begin{align}
\displaystyle
 \frac{\delta\nu_{n,l}}{\nu_{n,l}} = &
 \int K_{c,\rho}^{(n,l)} \, \frac{\delta_x c}{c} \dx
+
 \int K_{\rho,c}^{(n,l)} \, \frac{\delta_x (G \rho)}{G \rho} \dx
\nonumber \\
& \displaystyle \mbox{}
- \frac{\delta R}{R} \int K_{c,\rho}^{(n,l)} \dx
+
S_{n,l}
 \;.
\end{align}
This equation is identical to the following more compact expressions:
\begin{equation}
 \frac{\delta\nu_{n,l}}{\nu_{n,l}} = 
 \int K_{c,\rho}^{(n,l)} \, \frac{\delta_x (c/R)}{c/R} \dx
+
 \int K_{\rho,c}^{(n,l)} \, \frac{\delta_x (G \rho)}{G \rho} \dx
+
S_{n,l}
\end{equation}
and, {more pertinently,} 
\begin{equation}
\displaystyle
 \frac{\delta\nu_{n,l}}{\nu_{n,l}} = 
 \int K_{c,\rho}^{(n,l)} \, \frac{\delta_x \covr}{\covr} \dx
+
 \int K_{\rho,c}^{(n,l)} \, \frac{\delta_x (G \rho)}{G \rho} \dx
+
S_{n,l}
\;,
\label{eq:perturbed_VP_a}
\end{equation}
which {follows}  by noticing the relation,
\begin{equation}
\frac{\delta_x r}{r} = \frac{\delta_x (R x)}{R x} = \frac{\delta R}{R}
\;.
\label{eq:dxr_dR}
\end{equation}
{%
Equation (\ref{eq:perturbed_VP_a}) implies
$K_{\covr,\rho}^{(n,l)}=K_{c,\rho}^{(n,l)}$
and
$K_{\rho,\covr}^{(n,l)}=K_{\rho,c}^{(n,l)}$
in equation (\ref{eq:perturbed_VP})%
};  {neither contains} 
the radius difference $\delta R$ explicitly.

\section[]{Derivation of the annihilator relation}\label{sec:derive_ann}

%
{Associate with} 
a single structure two different scale factors $R$ and $R'$.
Let us call the structure with scale factor $R$ model A
and that with $R'$ model B.
The difference in the scale factors is given by
\begin{equation}
 \delta R := R' - R\;\neq 0.
\label{eq:B1}
\end{equation}
To relate the
radius $r$ in model A and radius $r'$ in model B
at the same fractional radius, we set
\begin{equation}
r'/R' = r/R 
\;,
\label{eq:B2}
\end{equation}
from which 
\begin{equation}
r' = \frac{R'}{R}\; r
   = \left(1 + \frac{\delta R}{R}\right) r
\;.
\label{eq:B3}
\end{equation}
Therefore
the difference 
between models A and B 
in any quantity $f$
at the same fractional radius $x$
is estimated as 
\begin{equation}
 \delta_x f 
 := f(r') - f(r)
 = \frac{\delta R}{R}\;\frac{{\rd} f}{{\rd}r}\;r
 = \frac{\delta R}{R}\;\frac{{\rd} f}{{\rd}\ln r}\;,
\label{eq:B4}
\end{equation}
which is valid in the linear regime.
Hence we have
\begin{equation}
 \frac{\delta_x f}{f} = \frac{\delta R}{R}\;\frac{{\rd}\ln f}{{\rd}\ln r}\;.
\label{eq:B5}
\end{equation}
Using this formula,
equation (\ref{eq:perturbed_VP}) reduces to
\begin{align}
\frac{\delta \nu_{nl}}{\nu_{nl}}
&=
\frac{\delta R}{R}
\left(
\int 
K_{\covr,\rho}^{(n,l)}\;
\frac{{\rd}\ln \covr}{{\rd}\ln r}
\dx
+
\int 
K_{\rho,\covr}^{(n,l)}\;
\frac{{\rd}\ln\rho}{{\rd}\ln r}
\dx
\right)
\nonumber \\ &\phantom{=} \mbox{}
+
S_{nl}
\;,
\label{eq:B6}
\end{align}
where
we have used
\begin{equation}
 \frac{{\rd}\ln(G\rho)}{{\rd}\ln r}
=
 \frac{{\rd}\ln\rho}{{\rd}\ln r}\;,
\label{eq:B7}
\end{equation}
because $G$ is constant.
Since 
the two structures are physically the same,
we necessarily have
$\delta\nu_{nl}/\nu_{nl}=0$, and
we
can ignore the surface term.
Furthermore, because  $\delta R/R\ne 0$,
we obtain
\begin{equation}
0
=
\int 
K_{\covr,\rho}^{(n,l)}\;
\frac{{\rd}\ln \covr}{{\rd}\ln r}
\dx
+
\int 
K_{\rho,\covr}^{(n,l)}\;
\frac{{\rd}\ln\rho}{{\rd}\ln r}
\dx
\;,
\label{eq:annihilator_b}
\end{equation}
for any $(n,\,l)$,
which is equation (\ref{eq:annihilator}).
{%
Strictly speaking,
we can show that
equation (\ref{eq:annihilator_b})
is correct only if
$K_{\rho,\covr}^{(n,l)}=0$
at the surface.
}


Like
equation (\ref{eq:perturbed_VP_a}),
equation (\ref{eq:annihilator_b})
is directly related to the homology relation
explained in the main text
that
adiabatic eigenfrequencies are invariant
under uniform stretching in the radial direction 
and appropriate scaling of the structure.


\section[]
{Inversion formulae 
and kernel identities
for
various inversion variables}
\label{sec:invform}

In the discussion in the body of the paper 
we have adopted
the sound speed $c$ and density $\rho$ 
for inversion variables.
In a
manner analogous 
to that
in
Appendices
\ref{sec:derive_pt_VP}
and
\ref{sec:derive_ann},
we can derive corresponding 
formulae
and
identities
for any other {seismologically} independent pair of 
structure variables.
The procedure is summarized in this subsection.
The
outcome is 
useful
not only
for {helioseismological} inversion, 
but also
when
we study
the {seismological} inversion for
the structure of other stars
\citep[cf.][]{Gough:1993ab,Takata:2002aa},
in which case 
we usually 
do not know
the precise values of the mass and radius.
The kernels
for variables other 
than $c$ and $\rho$ 
can be obtained
by the method
described by  
\citet{Gough:1996aa}.
For brevity 
we now omit
the surface term
in the inversion formulae

\subsection{Density and first adiabatic exponent}
If we choose the density $\rho$
and the first adiabatic exponent $\gamma_1$ 
as inversion variables,
we have
\begin{equation}
\frac{\delta\nu_{n,l}} {\nu_{n,l}}
=
\int K_{\rho, \gamma_1}^{(n,l)} \frac{\delta_x (G\rho)}{G \rho} \dx
+
\int K_{\gamma_1, \rho}^{(n,l)} \frac{\delta_x \gamma_1}{\gamma_1} \dx
\,.
\end{equation}
With this variable pair, terms explicitly containing 
the difference in mass or radius 
that might have arisen in the derivation of the kernels 
cancel out, as is the case in the derivation of equation (\ref{eq:perturbed_VP}). 
We need 
pay attention only to the fact
that
the difference between the reference model
and the target
should be
taken
not at the fixed absolute radius $r$
but
at fixed fractional radius $x$,
whose definition
implicitly
includes the radius $R$.
Essentially the 
same discussion 
as that in Appendix \ref{sec:derive_pt_VP}
leads us to the
identity,
\begin{equation}
 \int K_{\rho, \gamma_1}^{(n,l)} \dx
 = \frac{1}{2}
\;,
\label{eq:K_rho_gone_int}
\end{equation}
which corresponds
to
equation
(\ref{eq:it_identity});  
the annihilator relation,
\begin{equation}
\int K_{\rho, \gamma_1}^{(n,l)} 
\frac{{\rd}\ln \rho}{{\rd}\ln r} \dx
+
\int K_{\gamma_1, \rho}^{(n,l)}
 \frac{{\rd}\ln \gamma_1}{{\rd}\ln r} \dx
=
0
\;,
\end{equation}
is obtained
by the same scaling argument as that given 
in Appendix \ref{sec:derive_ann}.
{%
Equation
(\ref{eq:K_rho_gone_int})
is equivalent to
equation (14) of \cite{Reese:2012aa}.}

\subsection{Adiabatic sound-speed and
first adiabatic exponent}
\label{subsec:c_gone_pair}

When 
 $c^2$
and the first adiabatic exponent $\gamma_1$
are adopted as inversion variables,
the integral formula for the relative frequency perturbation is revised 
as follows:
\begin{align}
\frac{\delta\nu_{n,l}} {\nu_{n,l}}
&=
\int K_{c^2, \gamma_1}^{(n,l)} \frac{\delta_x \covr^2}{\covr^2} \dx
+
\int K_{\gamma_1, c^2}^{(n,l)} \frac{\delta_x \gamma_1}{\gamma_1} \dx
\nonumber\\
&\phantom{=}\mbox{}
+
\left(
\frac{1}{2}
-
\int
K_{c^2, \gamma_1}^{(n,l)}
\dx
\right)
\left(
\frac{\delta (GM)}{GM}
-
3\,
\frac{\delta R}{R}
\right)
\;,
\label{eq:VP_c2_g1}
\end{align}
which corresponds
to
equation (\ref{eq:dim_perturbed_VP_var}).
%
{%
In equation (\ref{eq:VP_c2_g1}), we have
used
$K_{\covr^2, \gamma_1}^{(n,l)}=K_{c^2, \gamma_1}^{(n,l)}$
and
$K_{\gamma_1, \covr^2}^{(n,l)}=K_{\gamma_1, c^2}^{(n,l)}$.
}
There are 
two new terms 
on the right-hand side
to take account of
the mass and radius differences.
Applying
the  scaling
discussed
in Appendix \ref{sec:derive_ann}
leads to the analogue of equation 
{(\ref{eq:annihilator_b})}%
:
\begin{equation}
 \int
K_{c^2, \gamma_1}^{(n,l)}
\left(
\frac{{\rd}\ln c^2}{{\rd}\ln r}
+
1
\right)
\dx
+
\int
K_{\gamma_1, c^2}^{(n,l)}
\frac{{\rd}\ln\gamma_1}{{\rd}\ln r}
\dx
=
\frac{3}{2}
\,.
\label{eq:ann_c2_g1}
\end{equation}
We
have only one kind of kernel identity
in this case
because
the coefficient
of the mass difference
does not vanish
in equation
(\ref{eq:VP_c2_g1}), 
unlike
the case 
in which
$c$ and $\rho$ 
are adopted as inversion variables.
{This is, in turn, because we need to
assume the total mass constraint to derive the expressions for $K_{c^2, \gamma_1}^{(n,l)}$ 
and $K_{\gamma_1, c^2}^{(n,l)}$
\citep[e.g.][]{Gough:1996aa}.}

\subsection{Isothermal 
{sound-speed}
and 
first 
{adiabatic-exponent}
}
\label{subsec:u_gone_pair}

{We consider the case
where the inversion variables are set to
the isothermal sound-speed, $u := p/\rho$,
and the first adiabatic exponent $\gamma_1$.}
Substituting
the
trivial relations,
$K_{c^2, \gamma_1}^{(n,l)}
=
K_{u, \gamma_1}^{(n,l)}$
and
$K_{\gamma_1, c^2}^{(n,l)}
=
K_{\gamma_1, u}^{(n,l)}
-
K_{u, \gamma_1}^{(n,l)}$,
into
equations
(\ref{eq:VP_c2_g1})
and
(\ref{eq:ann_c2_g1}),
yields 
\begin{align}
\frac{\delta\nu_{n,l}} {\nu_{n,l}}
&
=
\int K_{u, \gamma_1}^{(n,l)} \frac{\delta_x (u/r^2)}{u/r^2} \dx
+
\int K_{\gamma_1, u}^{(n,l)} \frac{\delta_x \gamma_1}{\gamma_1} \dx
\nonumber\\ &\phantom{=}\mbox{}
+
\left(
\frac{1}{2}
-
\int
K_{u, \gamma_1}^{(n,l)}
\dx
\right)
\left(
\frac{\delta (GM)}{GM}
-
3\,
\frac{\delta R}{R}
\right)
\end{align}
and
\begin{equation}
 \int
K_{u, \gamma_1}^{(n,l)}
\left(
\frac{{\rd}\ln u}{{\rd}\ln r}
+
1
\right)
\dx
+
\int
K_{\gamma_1, u}^{(n,l)}
\frac{{\rd}\ln\gamma_1}{{\rd}\ln r}
\dx
=
\frac{3}{2}
\;,
\end{equation}
respectively.

{%
We finally make a remark about the
difference between $c^2$ and $u$.
Since the analyses in Appendices
\ref{subsec:c_gone_pair}
and
\ref{subsec:u_gone_pair} are essentially the same,
it might possibly be accepted that
it makes little difference
whether we choose $c^2$ or $u$
as one of the inversion variables.
This is, however, not always true.
}
{As pointed out by \citet{1991wddogagkmjtLNP...388..111D}, for example, w}hen 
combined with the 
{helium mass fraction $Y$}, 
{$u$}
is
usually
preferred over $c$ or $c^2$ 
for the accompanying inversion variable
because it {avoids
putative divergences in the $c^2$ kernel in ionization zones of abundant elements  
at locations where 
$(\partial {\rm ln}  \gamma_1/ \partial {\rm ln} \rho)_{p,Y}=1$,
thereby obviating unnecessary mathematical complication.}

\bsp	
\label{lastpage}
\end{document}